\documentclass[12pt]{iopart}

\usepackage{iopams}
\usepackage{graphicx}
\usepackage{upgreek}
\usepackage{units}
\usepackage{braket}
\usepackage{verbatim}
\usepackage{color}
\usepackage{hyperref}
\newcommand{\ii}{\mathsf{i}}
\newcommand{\osix}{$^{16}$O$^+_2$}
\newcommand{\Mgtf}{$^{25}$Mg$^+$}
\newcommand{\Caf}{$^{40}$Ca$^+$}
\newcommand{\ee}{\mathrm{e}}
\newcommand{\state}{\ket{\Psi}}

\begin{document}

\title[Quantum logic spectroscopy of molecular oxygen ions]{Prospect for precision quantum logic spectroscopy of vibrational overtone transitions in molecular oxygen ions}

\author{Fabian Wolf, Jan C. Heip, Maximilian J. Zawierucha, Chunyan Shi}

\address{Physikalisch-Technische Bundesanstalt, Bundesallee 100, 38116 Braunschweig, Germany}
\author{Silke Ospelkaus}
\address{Institut f\"ur Quantenoptik, Leibniz Universit\"at Hannover, Welfengarten 1, 30167 Hannover}
\author{Piet O. Schmidt}
\address{Physikalisch-Technische Bundesanstalt, Bundesallee 100, 38116 Braunschweig, Germany}
\address{Institut f\"ur Quantenoptik, Leibniz Universit\"ut Hannover, Welfengarten 1, 30167 Hannover}
\ead{fabian.wolf@ptb.de}
\vspace{10pt}
\begin{indented}
\item[]
\end{indented}

\begin{abstract}
  Precision spectroscopy has been the driving force for progress of our physical understanding and still is a promising tool for the investigation of new physics.
  Molecules offer transitions which allow tests that are not possible in atomic systems.
  However, usually precision spectroscopy of molecules is challenging due to the lack of cycling transitions for state preparation and state detection.
  For molecular ions, this obstacle can be overcome by quantum logic spectroscopy, where dissipation for state preparation and detection is provided by a co-trapped atomic ion exploiting the shared eigenstates of motion.
  Here, we propose a full quantum logic spectroscopy scheme for molecular oxygen ions and theoretically investigate the feasibility of quantum logic-assisted state detection and preparation.
  Furthermore, we provide coupling rates for a direct single-photon quadrupole excitation of a vibrational overtone transition that can serve as a sensitive reference for tests of a possible variation of the proton-to-electron mass ratio.
\end{abstract}

\section{Introduction}
Recent progress in the control of single molecular ions via quantum logic~\cite{wolf_non-destructive_2016,chou_preparation_2017,sinhal_quantum-nondemolition_2020,chou_frequency-comb_2020, lin_quantum_2020, collopy_effects_2023}  opens the door for a novel class of high precision molecular spectroscopy experiments. In particular, the fact that molecules offer transitions that only involve a change in the state of the relative motion of the nuclei distinguishes them from atomic systems.
Combined with the control via quantum logic, which enabled the currently most accurate optical clock~\cite{brewer_27+_2019}, hybrid systems consisting of trapped atomic and molecular ions form a platform capable of achieving unprecedented accuracy for the spectroscopy of molecules.
Possible applications range from the search for new types of interactions that are not described by the standard model~\cite{safronova_search_2018}, for example fifth force tests~\cite{salumbides_bounds_2013}, to tests for a possible variation of fundamental constants, in particular the proton-to-electron mass ratio $\mu=m_{\mathrm{p}}/m_{\mathrm{e}}$~\cite{schiller_tests_2005,flambaum_enhanced_2007,hanneke_high_2016,kajita_accuracy_2017,kajita_test_2014}.
Current bounds on a possible variation of $\mu$ are derived from frequency comparisons between hyperfine states of cesium and an optical clock transition. In these experiments the sensitivity to $\mu$ is provided by the cesium clock.
However, in order to relate the hyperfine transition frequency in cesium to the proton-to-electron mass ratio, assumptions on the dependence of the proton mass to the magnetic moment of the nucleus have to be made.
Therefore, these tests cannot claim to be model independent~\cite{Flambaum_dependence_2006}.
The most stringent bounds using these assumptions are $\frac{\mathrm{d}}{\mathrm{d}t}\ln{\mu}=\unit[0.2(1.1)\times10^{-16}]{year^{-1}}$~\cite{godun_frequency_2014} and $\frac{\mathrm{d}}{\mathrm{d}t}\ln{\mu}=\unit[0.5(1.6)\times10^{-16}]{year^{-1}}$~\cite{huntemann_improved_2014}.
On the other hand, the dependence of molecular rotational and vibrational structure on $\mu$ relies on very basic principles that can be verified experimentally by isotope shift spectroscopy.
Therefore, these tests for variation of $\mu$ are often referred to as 'model-independent'.
The most stringent model-independent bound for a variation of $\mu$ is $\unit[0.3\pm1.0\times10^{-14}]{year^{-1}}$ and was set by spectroscopy on KRb-molecules~\cite{kobayashi_measurement_2019}.
An interesting candidate for improving this bound is the oxygen molecular ion~\cite{hanneke_high_2016,carollo_two-photon_2018,kajita_accuracy_2017}.
As a homonuclear molecule, it provides narrow transitions, since rotational and vibrational excitations are dipole forbidden.
Furthermore, the most abundant isotope $^{16}$O does not have nuclear spin, which simplifies the electronic level structure.
These features suggest the oxygen molecular ion also as an interesting candidate as a quantum memory for quantum information processing~\cite{mur-petit_toward_2013}.
However, control over the internal states of molecular ions is in general a challenge due to the lack of cycling transitions, which hinders state preparation and state detection.
As a consequence, previous spectroscopy of molecular ions was restricted in fractional precision to the $10^{-12}$ range for vibrational transitions~\cite{kortunov_protonelectron_2021,patra_protonelectron_2017} and to the $10^{-11}$ range for rotational transitions~\cite{alighanbari_precise_2020}.
A technique that eliminates these obstacles and has been used to push the achievable precision of rotational state spectroscopy to the $10^{-13}$ range~\cite{collopy_effects_2023} is quantum logic spectroscopy.
Quantum logic spectroscopy of molecular ions has been proposed already more than ten years ago~\cite{schmidt_spectroscopy_2006, vogelius_probabilistic_2006, leibfried_quantum_2012, ding_quantum_2012, mur-petit_temperature-independent_2012, shi_microwave_2013} and first implementations of quantum-logic assisted state detection~\cite{wolf_non-destructive_2016,chou_preparation_2017,sinhal_quantum-nondemolition_2020}, state preparation~\cite{chou_preparation_2017} and rotational spectroscopy~\cite{collopy_effects_2023} were reported recently.
The quantum logic approach relies on supporting the molecular spectroscopy ion with a co-trapped atomic logic ion.
The logic ion provides a transition for laser cooling and state manipulation, as well as state detection.
The Coulomb interaction strongly couples the individual motional modes to shared modes of motion, which allows reducing the kinetic energy of both ions by only applying cooling to the logic ion.
Furthermore, the shared motional modes can be used to transfer information on the internal state of the spectroscopy ion to the logic ion, where it can be read out efficiently.

Here, we propose a quantum logic protocol for spectroscopy of molecular oxygen ions that can be implemented with current state-of-the-art experimental setups.
The paper is structured as follows:
In section~\ref{sec:oxy} an overview of the relevant physical properties and electronic structure of molecular oxygen is given. This also includes an estimation of the sensitivity of overtone spectroscopy to a possible variation of $\mu$.
The next section deals with the proposed experimental sequence.
The following sections provide further details on different steps that are required for the proposed spectroscopic scheme.
Section~\ref{sec:Preparation} deals with the initial preparation of the two-ion crystal, section~\ref{sec:QuantumLogic} introduces the proposed quantum logic schemes for internal state preparation and detection and the interrogation of a vibrational overtone is discussed in section~\ref{sec:interrogation}.
In the last section the main findings are summarized.
\section{Level scheme and physical properties of $^{16}$O$_2^{+}$}\label{sec:oxy}
\begin{figure}[htp]
  \centering
  \includegraphics[width=\columnwidth]{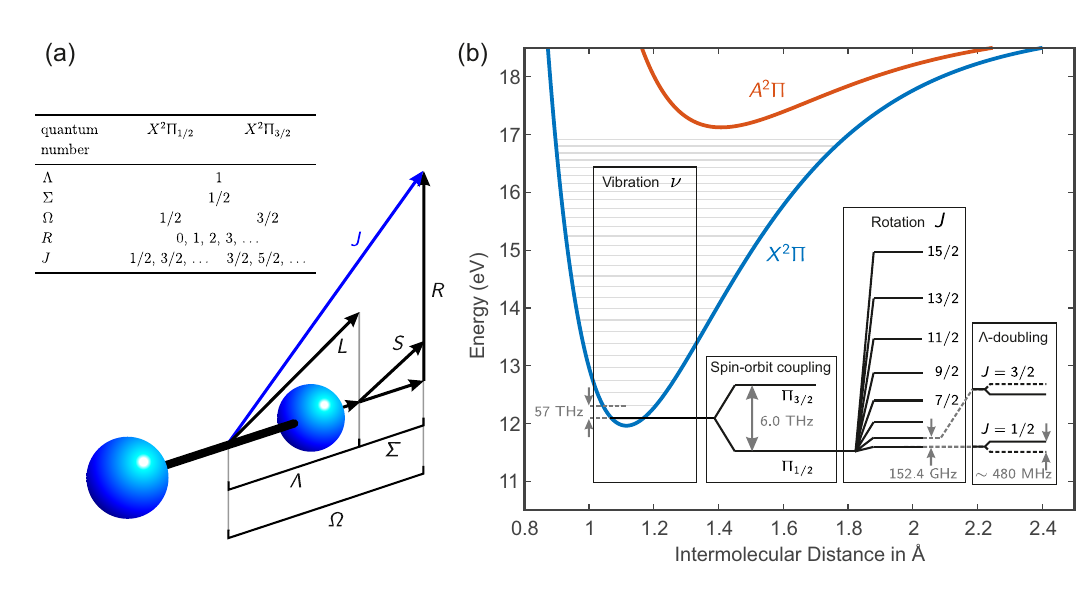}
  \caption{Energy levels of O$^+_2$. (a) shows the well-defined quantum numbers for the ground state of the oxygen molecular ion and the corresponding angular momentum coupling according to Hund's case a. (b) shows the reduced energy level diagram. The vibrational splitting and Morse potentials are taken from reference~\cite{carollo_two-photon_2018}, spin-orbit coupling constant, rotational splitting and $\Uplambda$-splitting are taken from reference~\cite{coxon_rotational_1984}.}
  \label{fig:LevelScheme}
\end{figure}

Oxygen naturally occurs in three different stable isotopes, $^{16}$O, $^{17}$O and $^{18}$O, with the relative abundances $\unit[99.759]{\%}$, $\unit[0.0374]{\%}$ and $\unit[0.2039]{\%}$~\cite{nier_redetermination_1950}, respectively.
Except for $^{17}$O~($I=5/2$), the stable isotopes have zero nuclear spin. Here, we mostly focus on the homonuclear ionic molecule of the most abundant isotope, \osix.
The X$^2\Uppi$ ground state of oxygen follows Hund's case (a) angular momentum coupling (see figure~\ref{fig:LevelScheme}~(a)). In consequence, the electron spin $S$ and orbital angular momentum $L$ are quantized with respect to the internuclear axis.
The molecular oxygen ion has a single unpaired electron, therefore the total spin in the electronic ground state is $S=1/2$ with a projection $\Sigma=\pm1/2$ along the internuclear axis. The projection of the electronic angular momentum on the internuclear axis is $\Lambda=1$. Spin-orbit interaction couples $\Sigma$ and $\Lambda$ to the total angular momentum $\Omega$, which results in two fine structure components $|\Omega|=1/2$ and $|\Omega|=3/2$ for the electronic ground state. The ambiguity in the sign of $\Omega$ gives rise to two degenerate states of opposite parity whose degeneracy is lifted by coupling to excited $\Sigma$-states which results in the so-called $\Lambda$-splitting. In the case of homonuclear oxygen molecules with vanishing nuclear spin, the nuclei have to follow Bose-Einstein statistics, therefore only states that are symmetric under inversion of the nuclei are allowed and the other half of the states are missing, making $\Lambda$-splitting only visible as a relative shift between the energy levels~\cite{coxon_rotational_1984}.
In terms of the molecular term symbols the ground states are labelled $X^2\Uppi_{|\Omega|}$. Apart from the electronic ground state, we consider only a single excited state $A^2\Uppi_{|\Omega|}$, which follows Hund's coupling case (b)~\cite{krupenie_spectrum_1972}. Coupling to other states is not considered here, because all other states are either energetically far separated from the ground state, or have a different multiplicity and therefore do not couple to the ground state via strong electric dipole transitions.

Figure~\ref{fig:LevelScheme}~(a) shows the relevant quantum numbers and subfigure (b) the reduced level scheme.
The energy eigenvalues for the $X^2\Uppi$ and $A^2\Uppi$ state can be inferred from the spectroscopic constants listed in table~\ref{specConst}.
The energy levels are expressed by~\cite{brown_rotational_2003}
\begin{equation}
  E/(hc)=T_e+T_\mathrm{rv}(J,\nu)+T_{\mathrm{so}}(\Omega,\nu)+T_{\Lambda}+T_{\mathrm{Zeeman}}(m_J)
  \label{eq:energylevels}
\end{equation}
where $T_e$ is the energy of the molecular potential minimum, $T_\mathrm{rv}$ is the ro-vibrational energy for total angular momentum quantum number $J$ and vibrational quantum number $\nu$,
$T_\mathrm{so}$ is the spin-orbit energy, that gives rise to the fine structure splitting.
The substructure due to Zeeman interaction with an external magnetic field is given by $T_\mathrm{Zeeman}$, where $m_J$ is the projection quantum number for the total angular momentum $J$ with respect to the magnetic field quantization axis.
$T_\Lambda$ is the energy shift from $\Lambda$-doubling.

The ro-vibrational energies can be expressed in terms of a Dunham expansion with Dunham coefficients $Y_{ij}$, which were experimentally determined for example by Prasad \textit{et al.} \cite{prasad_fourier_1997}. The values are summarized in table~\ref{specConst}. The corresponding ro-vibrational energy is given by
\begin{equation}
  T_\mathrm{rv}=\sum_{i,j}Y_{ij}\left( \nu+\frac{1}{2} \right)^i \left[ J\left( J+1 \right) \right]^j.
  \label{rovibDunham}
\end{equation}
In a similar fashion, the fine structure splitting can also be expressed in a Dunham-like expansion by
\begin{equation}
  T_\mathrm{so}=\sum_k X_k \left( \nu+\frac{1}{2} \right)^k.
  \label{soDunham}
\end{equation}

Experimental values for $X_k$ were determined by Coxon and Haley~\cite{coxon_rotational_1984}.
We would like to note that there are alternative sources for the ro-vibrational as well as the spin-orbit Dunham coefficients that provide either only theoretical values or slightly less accurate experimental values. An overview can be found in reference~\cite{liu_accurate_2015} and reference~\cite{song_rotationally_1999}. Most of the reported results agree with the data from Prasad~\textit{et al.}~\cite{prasad_fourier_1997} and Coxon and Haley~\cite{coxon_rotational_1984} within the experimental uncertainties.

The energy shift due to the $\Lambda$-doubling can be quantified by the $p$ and $q$ parameters and reads~\cite{hinkley__1972}
\begin{equation}
  T_{\Lambda}=\left(J+\frac{1}{2}\right)\left[ \left( \pm1-\frac{Y}{X}\pm \frac{2}{X}\right)\left( \frac{p}{2}+q \right)+\frac{2}{X} \left( J+\frac{3}{2} \right)\left( J-\frac{1}{2} \right)q \right]
  \label{eq:LambdaDoubling}
\end{equation}
for a $^2\Uppi$ state.
Here, $Y=A/B_\nu$ and $X^2=\left( Y\left( Y-4 \right)+4\left( J+\frac{1}{2} \right)^2 \right)$, where $A$ is the spin-orbit coupling constant and $B_\nu$ the rotational constant in the vibrational state $\nu$.

The Zeeman effect lifts the degeneracy of the different total angular momentum projection states, labelled by $m_J$. The corresponding energy shift for a magnetic field $B$ is given by~\cite{brown_rotational_2003} (see also \ref{app:Zeeman})
\begin{equation}
  T_\mathrm{Zeeman}=\left(\frac{\mu_B}{hc}\right)\frac{m_J}{J\left(J+1\right)} \left[\Omega\left(g_L\Lambda+g_s\Sigma\right)-g_r\left(J\left(J+1\right)-\Omega^2\right)\right]B
  \label{eq:ZeemanShift}
\end{equation}
with $g_L$, $g_S$ and $g_r$ the angular momentum, electron spin and rotational $g$-factor, respectively.
The $J$-dependence of the energy splitting between subsequent $m_J$ states due to the Zeeman shift will later be used to experimentally determine the $J$-state of the molecule.
In the following, we will use $g_L=1$ and $g_s=2.002$, which results in a Zeeman splitting of
\begin{equation}
  cT_\mathrm{Zeeman}(\Omega=3/2)=\unit[\frac{42}{J(J+1)}B]{MHz/mT}
  \label{eq:Zeeman32}
\end{equation}
and
\begin{equation}
  cT_\mathrm{Zeeman}(\Omega=1/2)=\unit[\frac{7}{J(J+1)}B]{kHz/mT}
  \label{eq:Zeeman12}
\end{equation}

Note, that we use the free electron's $g$-factor which might differ from the bound electron's g-factor in oxygen. Furthermore, we neglect effects from the rotational $g$-factor.
Therefore, in particular our estimates for the Zeeman splitting in the $\Omega=1/2$ fine structure state demands experimental verification or a more detailed theoretical investigation.
Theoretical values for the rotational Zeeman $g$-factor were only published for the ro-vibrational ground state in the $\Omega=1/2$ fine structure manifold ($g_\mathrm{r}(\Omega=J=1/2,\nu=0)=3.06\times10^{-5}$)~\cite{kajita_accuracy_2017}.

Figure~\ref{fig:LevelScheme}~(b) shows the reduced energy level scheme for the energetically lowest states and also provides orders of magnitudes for the involved energy splittings.

\begin{table}[h]
  \centering
  \caption{Spectroscopic constants of the Oxygen cation. $Y_{i,j}$ denotes the Dunham coefficients. The related label for the coefficient in the Morse potential is given in brakets. The values for the Dunham coefficients are from reference~\cite{prasad_fourier_1997} and were experimentally obtained via Fourier transform emission spectroscopy. $X$ are the Dunham-like parameters for the spin-orbit coupling from reference~\cite{coxon_rotational_1984}. All coefficient are given in $\unit{cm^{-1}}$}
\begin{tabular}{cccc}
\br
State  &  $X^2\Pi_g$  &  $A^2\Pi_u$  & reference \\\mr
$Y_{10}(\omega_e)$   &  $1905.892(82)$ & $898.65(12)$ & \cite{prasad_fourier_1997}\\
$-Y_{20}(\omega_ex_e)$ & $16.489(13)$ & $13.574(46)$ & \cite{prasad_fourier_1997}\\
$Y_{30}(\omega_ey_e)$ & $0.02057(90)$   & $-0.0066(51)$ & \cite{prasad_fourier_1997}\\
$Y_{40}(\omega_ez_e)$ & $-0.737(24)\times10^{-3}$ & & \cite{prasad_fourier_1997}\\
$Y_{01}(B_e)$ & $1.689 824(91)$ & $1.061939(14)$ & \cite{prasad_fourier_1997}\\
$-Y_{11}(\alpha_e)$ & $0.019363(37)$ & $0.019598 (16)$ & \cite{prasad_fourier_1997}\\
$Y_{21}(\gamma_e)$ & $-0.132(47)\times 10^{-4}$ & $-0.1019(30)\times 10^{3} $ & \cite{prasad_fourier_1997}\\
$Y_{31}$ & $-0.158(19)\times 10^{-5}$ &  & \cite{prasad_fourier_1997} \\
$X_{0}$ & $200.634(17)$ & & \cite{coxon_rotational_1984}\\
$X_{1}$ & $-0.6166(86)$ & & \cite{coxon_rotational_1984}\\
$X_{2}$ & $-6.94(140)\times 10^{-3}$ & & \cite{coxon_rotational_1984}\\
$X_3$ & $-7.01(70)\times10^{-4}$ & &\cite{coxon_rotational_1984}\\
\br
\end{tabular}
\label{specConst}
      \end{table}

\subsection{Sensitivity of ro-vibrational transitions to the proton-to-electron mass ratio}
\label{sec:sensitivity}
\begin{figure}[htpb]
  \centering
  \includegraphics[]{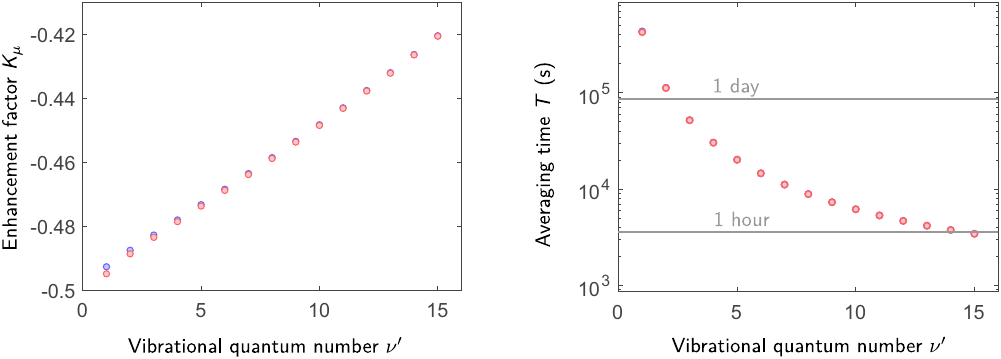}
  \caption{Figure of merit for sensitivity to a change in $\mu$. Left: enhancement factor $K_\mu$ for different vibrational overtone transitions $(\nu=0,J=1/2) \rightarrow (\nu',J')$. Blue circles correspond to $J'=3/2$ and red circles to $J'=5/2$. Right: minimum required averaging time for an overtone spectroscopy to infer a relative upper bound of  $\mathrm{d}\mu/\mu=10^{-16}$ on a possible variation of the proton-to-electron mass ratio with a Ramsey dark time of $\unit[300]{ms}$.}
  \label{fig:sensitivity}
\end{figure}

The sensitivity of ro-vibrational transitions to a possible variation of the proton-to-electron mass ratio can be estimated using the isotopic dependence of the Dunham coefficients $Y_{ij}$ on the reduced mass $M$ of the nuclei~\cite{brown_rotational_2003,watson_isotope_1980}.
This dependence can be approximated by
\begin{equation}
  Y_{i,j}\approx M^{-(i/2+j)}U_{ij}\approx\left(\frac{Z}{2}\mu \right)^{-(i/2+j)}\tilde{U}_{ij}
  \label{DunhamReducedMAss}
\end{equation}
with a nuclear mass-independent term $U_{ij}=\tilde U_{ij}m_\mathrm{e}^{i/2+j}$.
This expression only holds within the Born-Oppenheimer approximation. A more accurate description can be found in reference~\cite{watson_isotope_1980,tiemann_isotope_1982}.
In order to quantify an enhancement, we define the enhancement factor
\begin{equation}
  K_\mu=\frac{\mu_0}{\omega_0}\left. \frac{\partial \omega}{\partial \mu}\right|_{\mu=\mu_0}
    \label{eq:enhancementfactor}
\end{equation}
 such that
 \begin{equation}
   \frac{\mathrm{d}\omega}{\omega_0}=K_\mu \frac{\mathrm{d}\mu}{\mu_0} \,.
   \label{eq:enhancment2}
 \end{equation}
 We can infer the transition frequency for a ro-vibrational transition $(\nu,J)\rightarrow(\nu',J')$ from equation~\ref{rovibDunham} and find
\begin{eqnarray}
  \omega&=&\sum_{i,j} \left(\frac{Z}{2}\mu\right)^{-(i/2+j)} \tilde U_{ij}\\\nonumber
        &\times&\left\{ \left( \nu'+\frac{1}{2} \right)^i\left[ J'\left( J'+1 \right) \right]^j-\left( \nu+\frac{1}{2} \right)^i\left[ J\left( J+1 \right) \right]^j\right\}\,.
  \label{eq:enhancementFrequency}
\end{eqnarray}
Assuming that the initial state is $J=1/2$ and $\nu=0$ we can express the enhancement factor for an overtone transition, where $\Delta J=\pm 1$ and therefore $J'=3/2$ as
\begin{eqnarray}
  K_\mu(J=1/2,J'=3/2)=\\\nonumber
  -\frac{1}{\omega_0}\sum_{i,j}Y_{ij}2^{-(i+2j)}\left( \frac{i}{2}+j \right)\left( -3^j+15^j (1+2\nu')^i \right)
  \label{eq:enhancement3}
\end{eqnarray}
and for a transition, with $\Delta J=2$ and thus $J'=5/2$ as
\begin{eqnarray}
  K_\mu(J=1/2,J'=5/2)=\\\nonumber
  -\frac{1}{\omega_0}\sum_{i,j}Y_{ij}2^{-(i+2j)}\left( \frac{i}{2}+j \right)\left( -3^{j}+35^j (1+2\nu')^i \right)
  \label{eq:singlephotonenhancement}
\end{eqnarray}
It should be noted that the enhancement factor alone does not provide a good criterion to identify a suitable transition for measuring a possible variation of $\mu$. As can be seen from figure~\ref{fig:sensitivity}~(a), the absolute value of the enhancement factor decreases with increasing order of overtone. However, transitions with larger transition frequency $\omega_0$ provide a larger $Q$-factor and therefore improved statistical uncertainty $\sigma_y(\tau)= 1/(\omega_0 \sqrt{T_R \tau})$, where $\tau$ is the averaging time. Assuming a Fourier-limited interrogation with Ramsey dark time $T_R$, the time it takes to average quantum projection noise for a single molecule to a resolution of $\mathrm{d}\mu/\mu_0$ can be estimated by
\begin{equation}
  T\geq \frac{1}{\left[K_\mu\omega_0\left( \frac{\mathrm{d}\mu}{\mu_0}\right)\right]^2 T_R}
  \label{eq:TimeToAverage}
\end{equation}
Figure~\ref{fig:sensitivity}~(b) shows the minimum averaging time $T$ to resolve a change of $\mu$ on the order of $\mathrm{d}\mu/\mu=10^{-16}$ with an assumed interrogation time of $\tau=\unit[300]{ms}$.
The larger enhancement factor for lower order vibrational transitions is overcompensated by the loss in statistical uncertainty due to quantum projection noise, which suggests to aim for spectroscopy of higher overtones. We show later that especially the suppression of laser coupling for higher overtones shows an opposite trend and necessitates a compromise that will depend on the details of the experimental implementation.

\section{Experimental sequence}
The proposed setup involves a single molecular $^{16}$O$^{+}_{2}$ ion trapped together with an atomic logic ion in a linear Paul trap.
The logic ion provides a suitable optical transition for laser cooling and two long-lived states separated by an energy $E_\mathrm{qu}=\hbar\omega_\mathrm{qu}$.
These states form a qubit that can be used to store quantum information.
It can either be manipulated by laser or radio frequency interaction and can be read out using electron shelving~\cite{nagourney_shelved_1986}.
Due to the strong Coulomb repulsion between the ions, the eigenmodes of their motion involve movement of both ions and can therefore be considered using shared quantum state.
Typical trapping frequencies $\omega_\mathrm{m,i}$ are between a few hundred kilohertz and a few megahertz.
In the following, we will only consider a single motional mode along the axial direction which is sufficient for the proposed experiment and refer to its frequency as $\omega_\mathrm{m}$.
In the resolved sideband regime, quantum control over the motion can be obtained by applying laser pulses that are detuned from the qubit transition, coupling the internal and external degrees of freedom. These so-called red and blue sideband transitions excite the qubit and remove or add a quantum of motion, respectively.
They are addressed by tuning the laser such that it bridges an energy of $\hbar(\omega_{\mathrm{qu}}\pm\omega_\mathrm{m})$.
These control capabilities allow ground state cooling~\cite{wan_efficient_2015,rugango_sympathetic_2015,poulsen_sideband_2011} of the two-ion crystal and detection of motional excitation~\cite{monroe_resolved-sideband_1995,gebert_detection_2016,wolf_motional_2019,ohira_phonon-number-resolving_2019}.
In order to implement quantum logic routines between the atomic and the molecular ion an additional interaction is required, which couples the molecule's internal state to the motion. Here, we suggest coupling Zeeman states in the molecule to the motion by a far-detuned Raman laser~\cite{chou_preparation_2017} and implement a state dependent oscillating force (see section~\ref{sec:QuantumLogic} for further detail).
These features form the basic ingredient for the proposed experimental sequence.

\begin{figure}[htpb]
  \centering
  \includegraphics{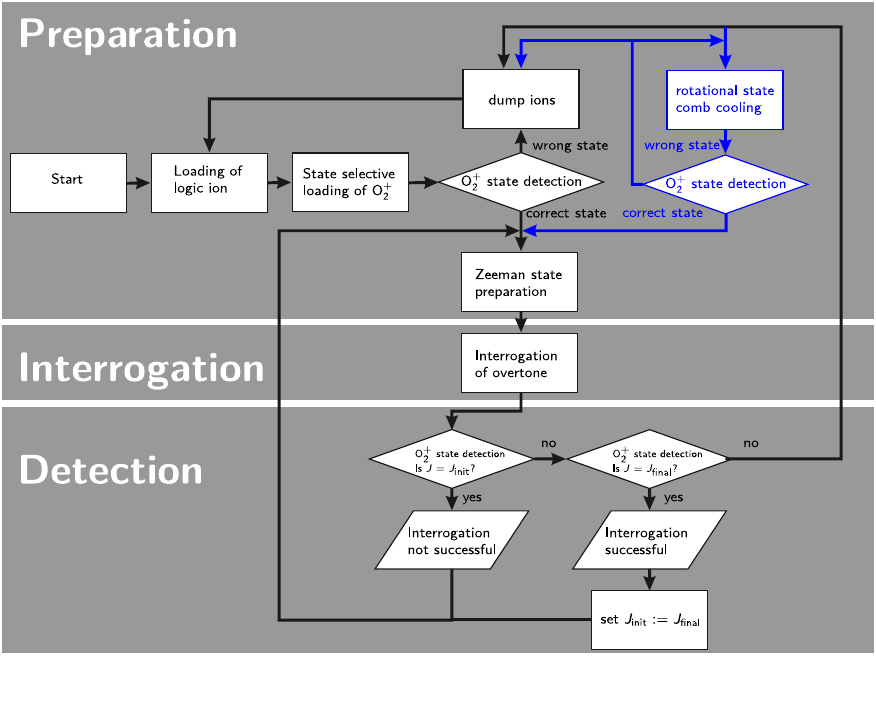}
  \caption{Flowchart for the experimental sequence. The flow chart with the black lines shows a sequence that does not require active rotational state preparation. The additional elements with blue lines show a modification using rotational state preparation with an optical frequency comb, which increases the duty cycle of the experiment by reducing the number of reloading sequences.}
  \label{fig:flowchart}
\end{figure}

A summary of the proposed experiment is shown in figure~\ref{fig:flowchart} in form of a flow chart.
After preparation of the two-ion crystal~(section~\ref{sec:Preparation}), the rotational and fine structure state of the \osix~ion is probed in a quantum-logic protocol~(see section~\ref{sec:detection}). If the ion is loaded with the wrong internal state, the two ion crystal is dumped and a new crystal is prepared.
If the oxygen ion is in the correct initial ro-vibrational state, the sequence proceeds with quantum logic-assisted preparation of the Zeeman state~(see section~\ref{sec:prep}), concluding initial state preparation for spectroscopy.
Different excitation schemes for interrogating the target transition are discussed in section~\ref{sec:interrogation}.
Finally, the internal state of the oxygen ion is again probed via quantum logic to determine if the interrogation was successful.
In case of successful depletion of the initial state, the population in the final state can also be checked by the quantum logic protocol as a cross check. Similarly, an excited initial state for spectroscopy can be prepared.

\section{Preparation of the two-ion crystal}\label{sec:Preparation}
The initial step of the experiment is the preparation of a two-ion Coulomb crystal and ro-vibrational state initialization of the oxygen molecular ion. We propose to prepare the molecular ion in the electronic and ro-vibrational ground state in the $\Omega=3/2$ fine structure manifold. The larger Zeeman shift in the $\Omega=3/2$ state simplifies state discrimination but also poses a challenge for precision spectroscopy due to its large electronic linear Zeeman shift. Therefore, it is advisable to switch to the $\Omega=1/2$ state for the final precision spectroscopy experiment when internal state manipulation and detection are well under control.
The techniques described in the following are applicable to both, the $\Omega=3/2$ and $\Omega=1/2$ manifolds.

First, the logic ion is loaded. Here, we consider well-developed logic ion species such as $^{25}$Mg$^+$ or $^{40}$Ca$^+$ due to the convenient charge-to-mass ratio with respect to \osix.
Lamb-Dicke Parameters and trapping frequencies for both combinations and convenient trap parameters are listed in table~\ref{tab:LambDickeParameter}.
Both logic ion species can be loaded via pulsed laser ablation loading, followed by resonant photo-ionization in the center of a linear Paul trap~\cite{hendricks_all-optical_2007,chen_ticking_2017,sheridan_all-optical_2011,hannig_towards_2019}. This process typically takes on the order of a few seconds.
Afterwards, the \osix~ion is loaded from a supersonic molecular beam~\cite{heip_ionization_2019} of neutral molecular oxygen.
The molecules in the beam are ionized in the trapping region by a 2+1 REMPI~(resonance enhanced multi-photon ionization) process using a pulsed UV Laser at around $\unit[300]{nm}$.
The resonant two-photon transition involved in the ionization process and the good Frank-Condon overlap between the excited Rydberg state and the ionic state allows to select the vibrational state of the molecular ion and restrict the number of possibly occupied angular momentum states~\cite{sur_optical_1985,dochain_production_2015,carollo_two-photon_2018}.

State-selective ionization of molecules~\cite{gardner_multi-photon_2019} and subsequent loading into an ion trap~\cite{tong_sympathetic_2010,tong_collisional_2011} has already successfully been demonstrated for nitrogen molecules. There, the final rotational state distribution was verified with light-induced charge transfer~(LICT).

To verify successful state-selective ionization, the next step in the initialization procedure is a non-destructive measurement of the $J$ state as described in subsection~\ref{sec:detection}.
Depending on the outcome of the state detection, the two-ion crystal is either dumped and the loading sequence starts again, or the experimental sequence proceeds with quantum logic-assisted Zeeman state pumping as described in the following section.

The duty cycle can be further increased by actively preparing the initial $J$ instead of dumping the molecules in unwanted $J$ states.  Schemes for active molecular state preparation using optical frequency comb-driven Raman sideband transitions between rotational states and dissipation via simultaneous sideband cooling on the logic ion have been proposed~\cite{ding_quantum_2012,leibfried_quantum_2012} and the demonstration of key techniques for their implementation have been reported recently~\cite{solaro_direct_2018,chou_frequency-comb_2020}. Other successfully implemented rotational state preparation schemes that rely either on nearly diagonal Franck-Condon factors~\cite{lien_broadband_2014}, vibrational state decay and blackbody radiation-induced rotational transitions~\cite{schneider_all-optical_2010,staanum_rotational_2010} are not applicable to oxygen.
Given the considerable time consumption associated with dumping and reloading ions compared to the spectroscopic sequence, it is highly desirable to minimize this process by, e.g., only dumping the molecular ion. This also underscores the urge for the development of efficient rotational state preparation schemes for molecular ions.
\begin{table}[h]
  \centering
  \caption{Logic ion specific trapping and coherent manipulation parameters. Lamb-Dicke parameter for quantum logic on the logic ion and in-phase~(IP) and out-of-phase~(OP) axial mode frequencies for $^{25}$Mg$^+$-\osix~and $^{40}$Ca$^+$-\osix~two-ion crystals. The trap parameters were chosen such that the trapping frequency for a single $^{25}$Mg$^+$ ion would be $\unit[2\pi\times 1.8]{MHz}$. The Raman laser beams for manipulation of \Mgtf were assumed to enclose an angle of $\unit[90]{^\circ}$ with an alignment that results in an effective wavenumber of $\Delta k=\sqrt{2}\times2\pi/\lambda$. For \Caf~it is assumed that the logic laser with wavelength of $\lambda=\unit[729]{nm}$ is aligned along the axial direction.}
  \begin{tabular}{lcccc}
    \br
    Logic ion species  & \multicolumn{2}{c}{Lamb-Dicke parameter} &\multicolumn{2}{c}{Mode frequency} \\
    & $\eta_{\mathrm{IP}}$ & $\eta_{\mathrm{OP}}$ & $\omega_{\mathrm{IP}}$  &$\omega_{\mathrm{OP}}$\\\mr     
   $^{25}$Mg$^+$ & $0.2146$  & $0.2068$ & $\unit[2\pi\times1.68]{MHz}$ & $\unit[2\pi\times2.95]{MHz}$\\
   $^{40}$Ca$^+$ & $0.0618$  & $0.0374$ & $\unit[2\pi\times1.50]{MHz}$ & $\unit[2\pi\times2.62]{MHz}$\\\br
  \end{tabular}

  \label{tab:LambDickeParameter}
\end{table}
\section{Quantum logic with molecular oxygen}\label{sec:QuantumLogic}
Having the \osix~ion trapped simultaneously with the atomic logic ion forms the starting point for quantum logic operations. Here, we propose to exploit quantum logic techniques for state preparation and state detection using a far-detuned continuous wave Raman laser setup as already demonstrated by Chou~\textit{et~al.}~\cite{chou_preparation_2017} for manipulation of $^{40}$CaH$^{+}$.
\subsection{Quantum logic using a far-detuned Raman laser}\label{sec:prep}
\begin{figure}[htpb]
  \centering
  \includegraphics{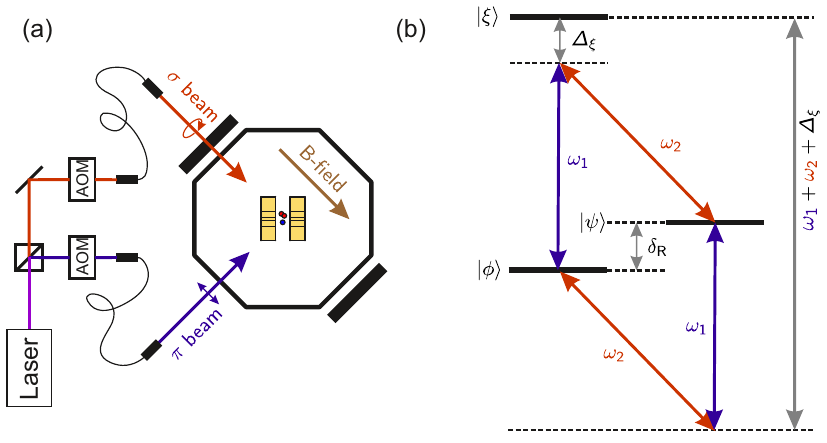}
  \caption{Laser setup for Raman transitions in \osix. (a) shows a sketch of the laser system and the ion trap in the vacuum chamber.
  The laser beam is split and shifted in frequency by acousto-optical modulators~(AOM) to obtain a relative detuning of $\omega_1-\omega_2=\delta_\mathrm{R}$. One laser beam carries $\pi$-polarization and the other $\sigma$-polarization. They enclose an angle of $90^\circ$ with a relative k-vector projection on the trap axis of $\Delta k=\sqrt{2}\times2\pi/\lambda$, where $\lambda$ is the wavelength of the laser light. (b) reduced level scheme for illustration of a Raman transition. The quantum states $\ket{\phi}$ and $\ket{\psi}$ are coupled via an excited state $\ket{\xi}$.  }
  \label{fig:RamanLaser}
\end{figure}

Two states, energetically separated by $\hbar \delta_\mathrm{R}$ can be coupled via an excited state in a two-photon Raman process.
The Raman transition is driven by two lasers with frequencies $\omega_1$ and $\omega_2$ and relative detuning of $\delta_\mathrm{R}=\omega_1-\omega_2$.
A detuning of the individual lasers with respect to the excited states suppresses off-resonant scattering, which is the major cause of decoherence in the process~\cite{ozeri_hyperfine_2005}. In molecular ions, spontaneous decay is particularly undesired since it is very likely to change the vibrational and rotational state of the ion.
A sketch of a laser setup and a reduced level scheme for a Raman transition is shown in figure~\ref{fig:RamanLaser}.
The Hamiltonian for Raman coupling between two states $\ket{\phi}$ and $\ket{\psi}$ via multiple excited states $\ket{\xi}$ is given by
\begin{eqnarray}
  _\psi\Omega_\phi^\mathrm{R}&=&\frac{1}{4\hbar}\sum_\xi \left(\frac{_\phi\Omega^{(1)}_\xi\, _\xi\Omega^{(2)}_\psi}{\Delta_\xi}+\frac{_\psi\Omega^{(1)}_\xi\, _\xi\Omega^{(2)}_\phi}{\omega_1+\omega_2+\Delta_\xi}\right)\,,
  \label{eq:RamanCoCount}
\end{eqnarray}
where $_\psi\Omega_\xi^{(1)}$ is the single-photon Rabi frequency between $\ket{\psi}$ and $\ket{\xi}$ (see \ref{subsec:DipoleTransitionMoment}).
The two terms in the sum correspond to the rotating and counter-rotating terms, which both need to be considered for large detuning.
For very large detuning $\Delta \gg \omega_{1,2}$ and a Raman coupling between two states of the same electronic-vibrational state $\tilde\phi$, expression~(\ref{eq:RamanCoCount}) can be rewritten as
\begin{eqnarray}
 _\psi\Omega_\phi^{R}&\approx\frac{1}{4\hbar}\sum_{\tilde\xi} \frac{|\mathcal{S}_{\mathrm{ev}(\tilde\phi,\tilde\xi)}|^2}{\omega_{\tilde\xi}}\\\nonumber
  &\times\underbrace{\sum_{\bar\xi}\left( \mathcal{S}^{(1)}_{\mathrm{rot}}(\bar\phi,\bar\xi) \mathcal{S}^{(2)}_{\mathrm{rot}}(\bar\xi,\bar\psi)+\mathcal{S}^{(2)}_{\mathrm{rot}}(\bar\psi,\bar\xi)\mathcal{S}^{(1)}_{\mathrm{rot}}(\bar\xi,\bar\phi)\right)}_{\mathcal{S}_{\mathrm{rot}}^{\mathrm{R}}}\,,
\end{eqnarray}
where we neglected the rotational splitting due to the large overall detuning and assumed no change in vibrational quantum number by going from $\phi$ to $\psi$ and therefore $\mathcal{S}_{\mathrm{ev}(\tilde\phi,\tilde\xi)}=\mathcal{S}_{\mathrm{ev}(\tilde\psi,\tilde\xi)}$.
$\mathcal{S}_{\mathrm{ev}}$ and $\mathcal{S}_\mathrm{rot}$ are the vibrational-electronic and rotational part of the single photon dipole Rabi frequency.
According to the Born-Oppenheimer approximation the molecular wave function $\ket{\phi}=\ket{\tilde\phi}\ket{\bar\phi}$ was separated into a radial $\ket{\tilde\phi}$ and an angular part $\ket{\bar\phi}$, and analogously for $\ket{\psi}=\ket{\tilde\psi}\ket{\bar\psi}$.
Further detail is given in \ref{sec:TransitionRates}.

Here, we consider coupling of neighbouring Zeeman states on the first order sideband transition, where each quantum added in the Zeeman manifold adds or removes a single quantum of motion.
Due to angular momentum conservation, this particular coupling requires a combination of a $\pi$- and a $\sigma$-polarized Raman beam.
Evaluating the rotational couplings in the Rabi frequency, it turns out that under the assumption that the splitting between the rotational P~($\Delta J=-1$), Q~($\Delta J=0$) and R~($\Delta J=1$) branch is not resolved, i.e. the detuning $\Delta_\xi$ is much larger than rotational splitting, coupling between Zeeman states $m_J=-1/2 \leftrightarrow m_J=1/2$ is suppressed.
Figure \ref{fig:RamanPumping} illustrates this for the example of the $J=3/2$ state.

\subsection{Preparation of Zeeman states}\label{subsec:prepZeeman}
\begin{figure}[htpb]
    \centering
    \includegraphics[width=\columnwidth]{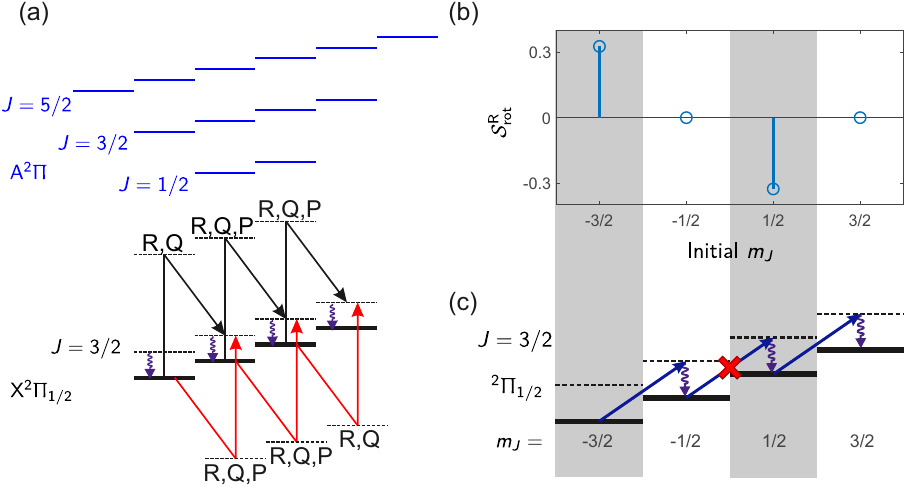}
\caption{(a) Schematic of the quantum logic-assisted pumping scheme.
A Raman configuration couples the different Zeeman states on the sideband transition.
The label (R,Q,P) denotes if the particular laser beam combination couples to the R, Q or P branch.
Black arrows represent the counter-rotating coupling, whereas red arrows denote co-rotating coupling.
The excited state follows Hund's case b.
Therefore, the sketched excited states are no eigenstates of the system.
For large detuning the structure is not resolved and an arbitrary basis can be chosen.
(b) Angular components of the couplings from the different branches assuming equal contributions from counter- and co-rotating terms.
It can be seen that the coupling that changes the sign of $m_J$ is suppressed as depicted in subfigure (c).  }
    \label{fig:RamanPumping}
  \end{figure}

For the preparation of the Zeeman state in O$^+_2$ we propose to implement a variant of the scheme demonstrated by Chou \textit{et al.}~\cite{chou_preparation_2017,schmidt_spectroscopy_2005}.
The initialization sequence starts with ground state cooling~\cite{wan_efficient_2015,rugango_sympathetic_2015,poulsen_sideband_2011} of the in-phase axial motional mode.
Then, a Raman sideband drive is applied that couples neighboring Zeeman states and adds a phonon of motion for each quantum added in the Zeeman degree of freedom.
Simultaneous sideband cooling on the logic ion provides a dissipation channel and breaks time reversal symmetry resulting in the molecule being pumped into a Zeeman edge state, as illustrated in figure~\ref{fig:RamanPumping}(a).
The Raman laser used for the Zeeman state preparation should be far off-resonant to avoid Raman scattering that would change the $J$ state~\cite{wolf_non-destructive_2016}.
For a given relative detuning $\delta_\mathrm{R}$, the direction of pumping is determined by the choice of polarization of the Raman beams ($\sigma$ and $\pi$).
In order to drive all transitions resonantly, relative ac-Stark shifts have to be suppressed, which can be achieved by choosing the intensity in the $\sigma$-polarized beam to be twice the intensity in the $\pi$-beam~\cite{chou_preparation_2017}.
Figure~\ref{fig:RamanPumping}(b) shows the angular part of the coupling rates between the individual $m_J$ states.
It shows that a change of the sign of $m_j$ is suppressed for the chosen type of coupling.
Depending on the sign of the initial $m_J$ state and the chosen polarization of the $\sigma$ Raman beam, the molecule is either pumped into the edge state $m_J=\pm J$ or into state $m_J=\pm 1/2$.
The state detection method, described in section~\ref{sec:detection} allows to distinguish these two states.
In case the ion is found in the wrong manifold for spectroscopy, resonant radio-frequency coupling can be used to invert the sign of $m_J$~\cite{shaniv_atomic_2016}.
Table~\ref{tab:LambDickeParameterFar} provides the Lamb-Dicke parameters for the axial motional modes for interaction with laser radiation at $\unit[532]{nm}$.
    \begin{table}[h]
  \centering
 \caption{Lamb-Dicke parameter for manipulation of the molecular ion in $^{25}$Mg$^+$-\osix~and $^{40}$Ca$^+$-\osix~two-ion crystals with $\lambda=\unit[532]{nm}$ laser light. The trap parameters were chosen such that the trapping frequency for a single $^{25}$Mg$^+$ ion would be $\unit[2\pi\times 1.8]{MHz}$. The Raman laser beams were assumed to enclose an angle of $\unit[90]{^\circ}$ with an alignment that results in an effective wavenumber of $\Delta k=\sqrt{2}\times2\pi/\lambda$.}

  \begin{tabular}{lccc}
    \br
    Logic ion species  & \multicolumn{2}{c}{Lamb-Dicke parameter}\\ 
    & $\eta_{\mathrm{IP}}$ & $\eta_{\mathrm{OP}}$ \\\mr     
    $^{25}$Mg$^+$ & $0.13$  & $0.08$ \\
   $^{40}$Ca$^+$ & $0.11$  & $0.10$ \\\br
  \end{tabular}
  \label{tab:LambDickeParameterFar}
\end{table}

\subsection{State detection}\label{sec:detection}
Extending the previous theoretical description of Raman coupling in the molecular ion, we propose in the following a new quantum logic scheme for the detection of the angular momentum state $J$ of the molecular ion directly after loading \osix~and after probing the spectroscopy transition~(see figure~\ref{fig:flowchart}).
It is based on resolving the state-dependent Zeeman splitting and therefore very similar to the scheme demonstrated by Chou \textit{et al.}~\cite{chou_preparation_2017}, where state-dependent splitting due to the coupling between the rotation and the nuclear spin was used in order to detect the rotational state.
However, we extend the scheme by suggesting a bichromatic drive, that allows to amplify the state detection signal which is in particular important if single-shot readout of the logic ion is technically not possible.
Similar to the optical pumping scheme, we propose to use a combination of a $\pi$-polarized and a $\sigma$-polarized Raman laser beam to drive transitions between neighboring Zeeman substates. As described previously, coupling of states with different sign of $m_J$ is suppressed. Therefore, the state $J=3/2$ is the first non-trivial state to be considered that can also be used to illustrate the detection mechanism.
\begin{figure}[htpb]
  \centering
  \includegraphics[width=89mm]{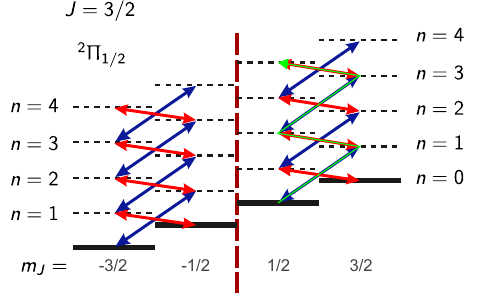}
  \caption{Schematic of the state detection by bichromatic Raman interaction. Shown are the $m_J$ magnetic substates of the $J=3/2$ rotational state in the electronic $^2\Uppi_{1/2}$ ground state. The dashed lines show the motional state ladder for each Zeeman state. Different Zeeman states are coupled along with an additional motional excitation or deexcitation by a blue sideband~(blue arrows) or red sideband~(red arrows), respectively. The simultaneous application of red and blue sideband transitions opens a path for motional excitation for each Zeeman component, which is illustrated on the example of the $m_J=1/2$ state by the green arrow. As described in the main text, a change of sign for $m_j$ is suppressed for far-detuned Raman lasers with the chosen polarization. }
  \label{fig:DetectionScheme}
\end{figure}
The only Raman coupling for $J=3/2$ is between $|m_J|=3/2$ and $|m_J|=1/2$, where the $m_J$ have the same sign.
Without loss of generality, we can therefore only consider the Zeeman substates with positive sign.
In consequence, the system is described by a qubit, $\ket{1/2}$ and $\ket{3/2}$.
After sideband cooling on the logic ion, we can use the far-detuned Raman laser to drive the red and blue sideband transitions between the Zeeman qubit states simultaneously.
In this scenario, the phase of the bichromatic drive can be chosen such that a Schr\"odinger cat state (see \ref{sec:SC}), $\ket{\psi}_\mathrm{SC}=\ket{+}\ket{\alpha}\pm\ket{-}\ket{-\alpha}$, is produced, where $\ket{\pm}=(\ket{1/2}\pm\ket{3/2})/\sqrt{2}$ and $\ket{\pm\alpha}$ denotes a coherent state with complex amplitude $\pm\alpha$.
The sign between the two parts of the wave function of the Schr\"odinger cat state is determined by the initial state.
The depletion of motional ground state due to the emergence of the Schr\"odinger cat state, indicates the successful drive of the transition and can be detected on the logic ion by implementing RAP on a red sideband transition~\cite{gebert_detection_2016,gebert_corrigendum_2018}.
Since the transition frequency depends on the molecule's $J$ state, the presence of motional excitation can be used as an indicator for the molecule's $J$ state.
An important feature of the scheme is that the motional ground state can be depleted irrespective of the initial Zeeman state, albeit not necessarily all the way to zero.

For $J>3/2$, the produced state is no longer an exact Schr\"odinger cat state.
In particular the fact that the coupling between neighboring $m_J$ states is not homogeneous, changes the motional dynamics.
Still, the bichromatic drive will lead to a reduction of the motional ground state population.
In order to avoid a small signal arising from the weakly coupled Zeeman components, quantum logic-assisted pumping to the edge states as described in \ref{subsec:prepZeeman} could be applied before detection.
Alternatively, the difference in maximum ground state depletion can be used as an indicator for the initial $m_J$ state.  Note that, for instance, in the $J=3/2$ manifold, the states $\ket{1/2}$ and $\ket{-3/2}$ exhibit identical Zeeman splittings and Rabi coupling rates to their neighboring states when only employing the far-detuned Raman coupling described here. As a result, they become indistinguishable. To differentiate between these states, the detection scheme would need to be modified, possibly incorporating additional couplings such as microwave fields.

\begin{figure}[htpb]
  \centering
  \includegraphics{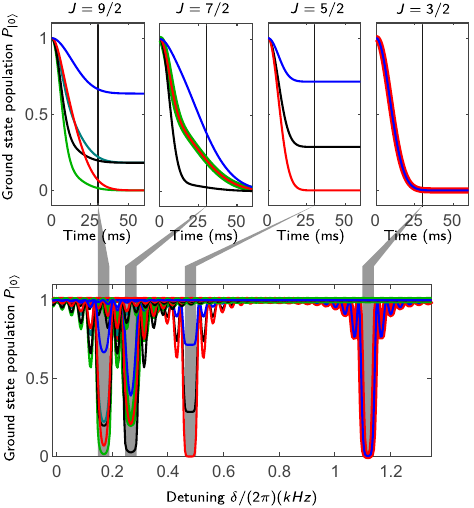}
  \caption{Simulation of the state detection protocol for the $\Omega=1/2$ manifold. The upper graphs show the depletion of the motional ground state for a resonant Raman interaction. The lower graph shows the spectrum of the state detection scheme. An individual rotational state can be detected by probing excitation at the corresponding detuning. The colors indicate the initial Zeeman state: $|m_J| = 1/2~\mathrm{(red)},\,3/2~\mathrm{(blue)},\,5/2~\mathrm({black)},\,7/2~\mathrm{(green)},\, 9/2~\mathrm{(teal)}$. The experimental parameters used for the simulation are: laser power $\sqrt{P_1P_2}=\unit[1]{W}$, beam waist $w=\unit[12]{\mu m}$, wavelength $\lambda=\unit[532]{nm}$, magnetic field $B=\unit[0.6]{mT}$.}
  \label{fig:Statedet12}
\end{figure}

\begin{figure}[htpb]
  \centering
  \includegraphics{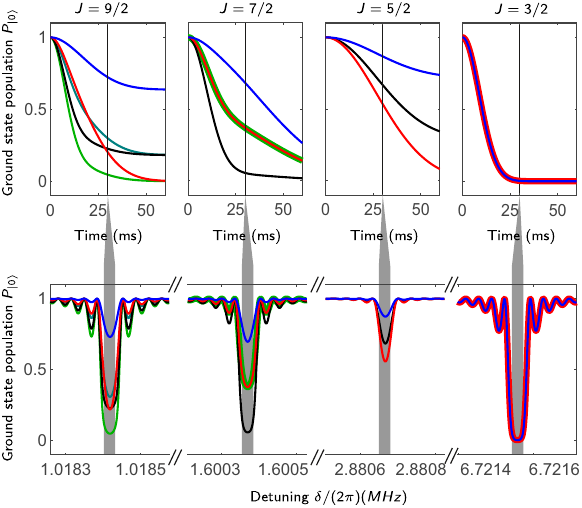}
  \caption{Simulation of the state detection protocol for the $\Omega=3/2$ manifold. The upper graphs show the depletion of the motional ground state for a resonant Raman interaction. The lower graph shows the spectrum of the state detection scheme. An individual rotational state can be detected by probing excitation at the corresponding detuning. The colors indicate the initial Zeeman state: $|m_J| = 1/2~\mathrm{(red)},\,3/2~\mathrm{(blue)},\,5/2~\mathrm{(black)},\,7/2~\mathrm{(green)},\, 9/2~\mathrm{(teal)}$. The experimental parameters used for the simulation are: laser power $\sqrt{P_1P_2}=\unit[1]{W}$, beam waist $w=\unit[12]{\mu m}$, wavelength $\lambda=\unit[532]{nm}$, magnetic field $B=\unit[0.6]{mT}$. }
  \label{fig:Statedet32}
\end{figure}

Figure~\ref{fig:Statedet12} and figure~\ref{fig:Statedet32} show the simulated spectrum using the QuTiP toolbox for python~\cite{johansson_qutip_2013} for a realistic experimental scenario.
The rotational Zeeman effect was neglected due to insufficient knowledge about the g-factor.
It can be seen that detection signal from neighbouring lines is easier resolved in $\Omega=3/2$ due to a larger Zeeman splitting. The rotational Zeeman shift is expected to be much smaller than the splitting between lines for different $J$.
Even in $\Omega=1/2$ the weak Rabi frequencies lead to such narrow lines that the peaks from different $J$-states are clearly separated.
However, the rotational Zeeman effect, that has been neglected here, might already be as large as the electronic Zeeman splitting and can potentially lead to overlapping lines.
We have assumed a Raman laser field with $\sqrt{P_1P_2}=\unit[1]{W}$ focused to a waist of $\unit[12]{\upmu m}$ with the Lamb-Dicke factor given in table~\ref{tab:LambDickeParameterFar} for Ca$^+$ as the logic ion.
For the chosen experimental parameters the detection time is in the tens of milliseconds range.
In an improved version of the detection scheme the required laser power and detection time can be substantially reduced by employing non-classical states, such as squeezed states~\cite{burd_quantum_2019}, Fock states~\cite{wolf_motional_2019} or Schr\"odinger cat states (where the motional state is entangled with the logic ion's internal state)~\cite{hempel_entanglement-enhanced_2013}, that allow quantum-enhanced displacement sensing.

  \section{Interrogation}\label{sec:interrogation}
For homonuclear species the excitation of vibrational transitions is dipole-forbidden for single photons. However, driving the transitions with multi-photon or higher multipole excitations is possible.
Recently, Carollo \textit{et al.}~\cite{carollo_two-photon_2018} proposed to excite a vibrational overtone transition in oxygen with two photons from the same laser field. Here, we discuss an alternative approach, namely the single-photon quadrupole excitation and compare it to a two-photon dipole excitation.
\subsection{Single photon quadrupole excitation}
Direct excitation of a dipole-forbidden vibrational transition in a molecular ion has been demonstrated in N$^+_2$ by Germann \textit{et al.}~\cite{germann_observation_2014}. They have performed spectroscopy on three components of the $\nu=0$ to $\nu'=1$ transition with a quantum cascade laser in the mid-infrared and detected loss of ions after state selective charge exchange reactions~\cite{tong_sympathetic_2010}.
Transition rates for direct quadrupole excitation of different vibrational overtones, starting from $\nu=0$ are listed in table~\ref{tab:quad}, for a spectroscopy laser with power of $\unit[1]{W}$ focussed to a waist of $w=\unit[10]{\upmu m}$.
It was assumed that the transition was driven between $m_J=1/2$ and $m_J'= 3/2$ with $\sigma^+$-polarized light and alignment of the laser along the quantization axis set by the magnetic field, which provides the largest transition rates.
Transition rates for other states, polarizations and laser orientations are given in the appendix in figures~\ref{fig:quadEx32} and \ref{fig:quadEx52}.
In section~\ref{sec:sensitivity} it was shown that higher order overtone transitions are better suited for a test of a possible variation of $m_\mathrm{p}/m_{\mathrm{e}}$ due to the larger involved energy splitting and the resulting reduction of statistical uncertainty. However, the transition rates drop drastically with increasing overtone order and reach the level of tens of Hz already for the $\nu=0$ to $\nu'=4$ transition. This renders spectroscopy challenging considering the current imprecise knowledge of the transition frequencies. 
It should be noted that the discretization of the used wavefunctions from reference \cite{carollo_two-photon_2018} introduce significant errors in the determination of the transition probabilities, therefore the given values should not be regarded as precise predictions but rather as approximate estimations of the transition probability. 
The transitions with $\nu'<3$ show reasonable transition rates for a broadband search of the transition frequency.
However, for the corresponding wavelength range above $\unit[2]{\upmu m}$ lasers with sufficient power are a technical challenge.
Commercially available quantum cascade lasers, optical parametric oscillators, difference frequency generation or Cr:ZnSe lasers are possible sources for coherent spectroscopy light.

\begin{table}
\centering
\caption{Maximum quadrupole excitation rates $\Omega$ in Hz  for low-lying vibrational states in the electronic ground state $X^2\Pi_g$. Transition: $J = 1/2, m_J = 1/2 \rightarrow J^\prime = 5/2, m_J = 3/2$. Laser power: $1\,$W; waist: $10\,\mu$m. The transition frequencies are taken from reference \cite{carollo_two-photon_2018}. }
\begin{tabular}{cccc}
\br
$\nu \rightarrow \nu'$ & $\lambda$  & $\Omega/2\pi$  & $\Omega/(2\pi \sqrt{I_\mathrm{laser}})$ \\
&($\upmu$m)&(Hz)&(Hz/$\sqrt{W/m^2}$)\\\mr
$0 \rightarrow 1$ & $5.307$  & $6.62\times10^4$ & $8.30\times 10^{-1}$ \\
$0 \rightarrow 2$ & $2.693$  & $4.31\times10^3$ & $5.40\times 10^{-2}$ \\
$0 \rightarrow 3$ & $1.8085$ & $4.97\times10^2$ & $6.23 \times 10^{-3}$\\
$0 \rightarrow 4$ & $1.369$  & $14.87$          & $1.86\times 10^{ -4}$\\
$0 \rightarrow 5$ & $1.1055$ & $41.42$          & $5.19 \times 10^{-4}$\\
$0 \rightarrow 6$ & $0.9295$ & $20.28$          & $2.54 \times 10^{-4}$\\
\br
\end{tabular}
\label{tab:quad}
\end{table}

\begin{figure}[htpb]
  \centering
  \includegraphics{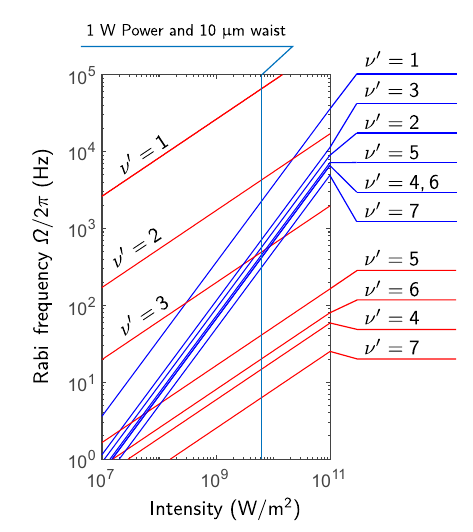}
  \caption{Comparison of excitation rates for two-photon dipole~(blue lines) and single-photon quadrupole excitation~(red lines) for vibrational (overtone) transitions with different final states $\nu'$. The intensity at the vertical blue line corresponds to a laser beam with $\unit[1]{W}$ power focussed to a waist of $\unit[10]{\upmu m}$.  }
  \label{fig:HannekeComp}
\end{figure}
\subsection{Comparison to other excitation schemes}
Carollo \textit{et al.}~\cite{carollo_two-photon_2018} proposed to excite a vibrational overtone in \osix~by a two photon dipole transition.
A major difference to the single photon quadrupole excitation scheme described here is the different scaling of the excitation rate $\Omega$ with the laser intensity $I$.
The quadrupole transition rate is proportional to $\sqrt{I}$, whereas the two-photon transition scales linearly with $I$.
Therefore, for each vibrational overtone a threshold intensity exists, where the coupling via the two-photon transition becomes stronger than the quadrupole coupling.
Figure~\ref{fig:HannekeComp} shows the coupling strengths in dependence of the laser intensity for different vibrational overtones for both, the quadrupole and the two-photon transition.
For the comparison, we have chosen $(J=1/2,m_J=1/2)$ as the initial state and $(J=5/2,m_J=3/2)$~(quadrupole) and $(J=5/2,m_J=-3/2)$~(two-photon dipole) as the final state.
These transitions provide the largest angular state couplings for the respective excitation schemes.
The electronic and vibrational transition strength for the two-photon dipole transition was taken from reference~\cite{carollo_two-photon_2018} and complemented by the angular component of the transition moment, which is similar to the angular part of the transition moment for a Raman transition, that is derived in \ref{sec:Raman}.
From the comparison, we can see that Rabi frequencies on the order of hundreds of hertz, which we consider sufficiently large for a broadband search of the transition frequency, require less power on the quadrupole transition for $\nu'\lesssim3$ and less power on the two-photon transition for $\nu'\gtrsim3$, depending on the available laser power.
However, for precision Rabi spectroscopy, where low Rabi frequencies on the order of a few hertz are required the quadrupole transition requires less power than the two-photon transition, which is advantageous to avoid light shifts.

Apart from the previously discussed excitation schemes, there are more exotic variants such as Raman transitions.

Raman transitions between vibrational transitions are a common tool for vibrational state manipulation~\cite{ni_high_2008,ospelkaus_efficient_2008} in neutral molecules and have been demonstrated as a tool for precision spectroscopy of vibrational transitions~\cite{kondov_molecular_2019}.
Instead of driving the vibrational transition with two photons of the same frequency as proposed by Carollo \textit{et al.}~\cite{carollo_two-photon_2018}, Raman transitions rely on two photons with different frequencies $\omega_1$ and $\omega_2$. The difference frequency $\delta=\omega_2-\omega_1$ has to match the vibrational spacing $\omega_\nu$.
In comparison to a two-photon drive with $\omega_1+\omega_2=\omega_\nu$, the scheme has the advantage, that near-resonant coupling to the electronically excited state A$^2\Uppi$ can enhance the coupling by orders of magnitude by not only exploiting the static but also the dynamic polarizability of the involved states.
However, in \osix~ laser radiation of around $\unit[200]{nm}$ would be required to benefit from near resonant coupling.
Moreover, a small detuning would also result in enhanced off-resonant scattering that would remove the molecule from the initial quantum state and therefore from the spectroscopic cycle.
We would like to note that our framework for calculating the transition strength for Raman transitions with far-detuned laser beams would need to be modified to take into account the change of angular momentum coupling between the X$^2\Uppi$ and A$^2\Uppi$ states, if the detuning is reduced to below the splitting between the rotational states.

An alternative to the near resonant Raman drive is a far off-resonant Raman laser setup.
Compared to a two-photon single frequency excitation, this approach offers the possibility to engineer a larger Lamb-Dicke factor, control ac-Stark shifts by tuning the polarization and achieve more flexibility in the choice of laser system at the expense of a more complex laser setup, involving phase locking the two Raman arms via a frequency comb.

\section{Conclusion and discussion}
In summary, we have proposed an experimental sequence for quantum logic spectroscopy of molecular $^{16}$O$^+_2$ ions.
The proposal addresses both issues caused by the lack of cycling transitions, namely state preparation and state detection with quantum logic techniques.
We have shown that despite the different electronic structure, the state preparation scheme that was developed and experimentally realized with CaH$^+$ by Chou \textit{et al.}~\cite{chou_preparation_2017} can be applied to oxygen in a modified way.
Complementary to the proposal by Carollo \textit{et al.}~\cite{carollo_two-photon_2018}, where a two photon drive was suggested we have evaluated the direct single photon quadrupole excitation strength for vibrational overtones including also the angular part of the wave functions.
Finally, we have theoretically developed a quantum logic-assisted state detection scheme that relies on resolving the rotational-state dependent Zeeman splitting and is therefore not only applicable to the oxygen molecular ion's ground state.
The detection scheme could also be applied to other rotational state dependent energy splittings such as the nuclear spin rotation coupling for example in $^{25}$MgH$^+$ or $^{40}$CaH$^+$.

We would like to note that a successful implementation of quantum logic spectroscopy on molecular oxygen requires further experimental investigations, some of which we discuss in the following.
Long averaging times of the overtone's transition frequency require an efficient ro-vibrational state-selective ionization process, which is currently under investigation in our group.
Furthermore, inelastic rotational state-changing collisions between \osix~and residual background gas~\cite{dorfler_rotational-state-changing_2020} can limit the available spectroscopy time.
In case of strong mixing of rotational states by collisions, either technical measures to improve the vacuum conditions have to be taken, or active rotational state preparation as described in references~\cite{ding_quantum_2012,leibfried_quantum_2012} has to be implemented.
Also, excessive motional heating in the ion trap can hinder the state detection and preparation.
Unwanted motional heating should be well below the anticipated motional excitation in the detection process which is on the order of $\frac{\unit[1]{phonons}}{\unit[25]{ms}}(=\unit[40]{phonons/s})$ which is frequently achieved in macroscopic ion traps even at room temperature.
Once the vibrational overtone spectroscopy is operational, a detailed systematic shift analysis needs to be performed, verifying the theoretically predicted uncertainty~\cite{kajita_accuracy_2017,carollo_two-photon_2018}.
Repeated frequency comparisons with optical clocks using an optical frequency comb~\cite{stenger_ultraprecise_2002,kajita_evaluation_2016} will then also allow establishing upper bounds on a possible variation of the fine structure constant and the proton-to-electron mass ratio~\cite{schiller_tests_2005,flambaum_enhanced_2007,hanneke_high_2016,kajita_accuracy_2017,kajita_test_2014}.

\section*{Acknowledgments}
We thank E. Tiemann and M. Kajita for valuable discussions. This work has been funded by the Deutsche Forschungsgemeinschaft (DFG, German Research Foundation) through CRC 1227 (DQ-mat), projects A04 and B05 with partial support from Germany^^e2^^80^^99s Excellence Strategy ^^e2^^80^^93 EXC-2123 QuantumFrontiers ^^e2^^80^^93 390837967.
\appendix
\section{Spherical tensor algebra}
In contrast to atomic systems, the internuclear axis in molecules provides an intrinsic quantization axis. Here, we consider a Hund's case (a) molecule. Consequently, the projection of the electron orbital angular momentum and spin on the molecular axis are defined. In order to evaluate coupling to fields with fixed polarization in the laboratory-fixed frame, a basis transformation has to be performed. Spherical tensor notation provides a handy tool for this task.

Following the notation from reference~\cite{brown_rotational_2003} we denote a spherical tensor of rank $k$ for an operator $A$ as $T^k(A)$. For an electric field $E$, the connection to cartesian coordinates (x,y,z) is given by
\begin{eqnarray}
  T^1_0(E)&=&E_z\,,\\
  T^1_{-1}(E)&=&-\frac{1}{\sqrt{2}}(E_x+\ii E_y)\,,\\
  T^1_{1}(E)&=&\frac{1}{\sqrt{2}}(E_x-\ii E_y)\,.\\
  \label{eq:sphericalTensor}
\end{eqnarray}
Therefore, the components $T^1_0$,$T^1_{-1}$ and $T^1_1$ of the laser field correspond to $\pi$, $\sigma_-$ and $\sigma_+$ polarization.
For an electric field gradient of a plane wave it is
\begin{eqnarray}
T^{(2)}_0\left(\nabla\vec{E}\right) &=& \sqrt{\frac{3}{2}}k_z E_z \\
T^{(2)}_{\pm 1}\left(\nabla\vec{E}\right) &=& \mp \frac{1}{2}\left[k_z E_x + k_x E_z \pm i \left(k_z E_y + k_y E_z\right)\right]\\
T^{(2)}_{\pm 2}\left(\nabla\vec{E}\right) &=& \frac{1}{2}\left[k_xE_x - k_yE_y \pm i\left(k_xE_y+k_yE_x\right)\right]
\end{eqnarray}
In the following, we will denote a spherical tensor component in the laboratory frame and the molecule fixed frame with a $p$ or $q$ subscript, respectively.
A spherical tensor in the laboratory fixed frame $T^k_p(A)$ can be transformed to the molecular fixed frame using the $k^\mathrm{th}$ rank Wigner rotation matrix $D^{(k)}_{pq}(\omega)$, with the Euler angles $\omega=(\phi,\theta,\chi)$ by
\begin{equation}
  T^k_p(A)=\sum_q D^{(k)}_{pq}(\omega)^* T^k_q(A)
  \label{eq:SphericalTensorTransformation}
\end{equation}
The matrix elements of the Wigner rotation matrix in the basis of the angular molecular wavefunction $\ket{\bar{\phi}}=\ket{\Omega,J,m_J}$ can be expressed as
\begin{eqnarray}
  \bra{\Omega,J,m_J}D^{(k)}_{pq}(\omega)^*\ket{\Omega',J',m_J'}=(-1)^{m_J-\Omega}\sqrt{\left( 2J'+1 \right)\left( 2J+1 \right)}\\\nonumber
  \times
  \left(
  \begin{array}{ccc}
    J & k  & J' \\
    -\Omega & q & \Omega' 
  \end{array}
\right)
  \left(
  \begin{array}{ccc}
    J & k  & J' \\
    -m_J & p & m_J' 
  \end{array}
\right).
  \label{eq:WignerExp}
\end{eqnarray}

\section{Zeeman interaction}\label{app:Zeeman}
The Zeeman Hamiltonian is given by~\cite{brown_rotational_2003}
\begin{eqnarray}
  H_\mathrm{Z}&=& \mu_\mathrm{B} \left( g_L+g_r\right)T^1(B)\cdot T^1(L)+\mu_\mathrm{B}\left( g_s+g_r \right)T^1(B)\cdot T^1(S)\\\nonumber
  &-&\mu_\mathrm{B}g_r T^1(B)\cdot T^1(J),
  \label{eq:ZeemanHamiltonian}
\end{eqnarray}
with the magnetic field $B$, the electronic orbital angular momentum $L$, the electron spin $S$ and the total electronic angular momentum $J$.
Since the electron's orbital angular momentum and spin are quantized in the molecule-fixed frame, their projections have to be rotated by Wigner rotation matrices in order to evaluate the matrix elements.
The laboratory coordinate system is chosen such that the magnetic field is aligned with the laboratory fixed z-axis. For the first term of $\bra{\bar\phi}H_\mathrm{Z}\ket{\bar\phi}$ we find
\begin{eqnarray}
  \bra{\bar\phi}T^1_{p=0}(B)\cdot T^1_{p=0}(L)\ket{\bar\phi}&=&B_z\Bra{\bar\phi}\sum_q D_{pq}^{(1)}(\omega)^*T_q^1(L)\Ket{\bar\phi}\\
  &=& B_z \Lambda \Bra{\bar\phi}D^{(1)}_{00}(\omega)^*\Ket{\bar\phi},
\label{LBExp}
\end{eqnarray}
where $B_z$ is the magnetic field component in z-direction in the laboratory frame.
We can use Eq.~\ref{eq:WignerExp} with $k=1$ and $q=p=0$
\begin{equation}
  \bra{\bar{\phi}}D^{(1)}_{00}(\omega)^*\ket{\bar{\phi}}=\frac{m_J\Omega}{J(J+1)}
  \label{eq:WignerExp00}
\end{equation}
and find
\begin{equation}
  \Bra{\bar\phi}T^1_{p=0}(B)\cdot T^1_{p=0}(L)\Ket{\bar\phi}=\frac{B_z\Lambda\Omega m_J}{J(J+1)}
  \end{equation}
The electronic spin Zeeman effect can be handled similarly, while $J$ is quantized with respect to the laboratory fixed magnetic field
\begin{equation}
  \bra{\bar\phi}T^1_{p=0}(B)\cdot T^1_{p=0}(J)\ket{\bar\phi}=Bm_J
\end{equation}
Hence, the Zeeman energy shift can be written as
\begin{eqnarray}
  E_\mathrm{Zeeman}&=&\bra{\bar{\phi}}H_\mathrm{Z}\ket{\bar{\phi}}\\
  &=&\frac{\Omega m_J B \mu_\mathrm{B}}{J(J+1)}\left[ \left( g_L+g_r \right)\Lambda+\left( g_s+g_r \right)\Sigma-g_r \frac{J(J+1)}{\Omega}  \right]\\
  &=&\mu_\mathrm{B}\frac{m_J}{J(J+1)}\left[\Omega\left( g_L\Lambda +g_s\Sigma \right)-g_R\left( J(J+1)-\Omega^2 \right)\right]
\end{eqnarray}

\section{Transition rates}\label{sec:TransitionRates}
\subsection{Dipole transition moment}\label{subsec:DipoleTransitionMoment}
The single photon electric dipole coupling between the states $\ket{\psi}$ and $\ket{\phi}$ induced by a laser can be described by the resonant Rabi frequency
\begin{equation}
_\psi\Omega_\phi^\mathrm{d}=\sum_p\frac{\braket{\psi|T^1_p(E)\cdot T^1_p(d)|\phi}}{\hbar}
  \label{psiomegaphi}
\end{equation}
where $T^1_p(E)=|E|T^1_p(\epsilon)$ and $T^1_p(d)=|d|T^1_p(\delta)$ are first rank spherical tensors, describing the electric field of the laser and the dipole moment in the laboratory frame~(denoted by index $p$), respectively.
$|E|$ and $|d|$ give the magnitude of the respective quantity and the orientation information is contained in the spherical tensors $T^1_p(\epsilon)$ and $T^1_p(\delta)$. 
According to the Born-Oppenheimer approximation, we can separate the radial part of the wave function $\tilde\psi(R)$ from the angular part $\bar\psi(\theta,\Phi)$ and find that the Rabi frequency can be written as a product of an electronic-vibrational $\mathcal{S}_\mathrm{ev}$ and a rotational part $\mathcal{S}_\mathrm{rot}$
\begin{equation}
  _\psi \Omega_\phi^\mathrm{d} = \frac{1}{\hbar} \underbrace{\Bra{\tilde \psi}\left(|d(R)||E|\right)\Ket{\tilde \phi}}_{\mathcal{S}_\mathrm{ev}}\underbrace{\left(\sum_p\Bra{\bar \psi}T^1_p(\epsilon)\cdot T^1_p(\delta)\Ket{\bar \phi}\right)}_{\mathcal{S}_\mathrm{rot}}\,.
  \label{omegaproduct}
\end{equation}
The orientation of the dipole moment $T^1_p(\delta)$ is fixed in the molecular frame (labelled by index $q$ in the spherical tensor notation), therefore it is convenient to write
\begin{eqnarray}
  \mathcal{S}_\mathrm{rot}(\bar{\psi},\bar{\phi})&=&\sum_p\bra{\bar{\psi}}T_p^1(\epsilon)\sum_q D_{pq}^{(1)}(\omega)^* T^1_q(\delta)\ket{\bar{\phi}} \\
  &=& \sum_p \epsilon_p\bra{\bar{\psi}}D_{p0}^{(1)}(\omega)^*\ket{\bar{\phi}}\,.
  \label{SrotSixJ}
\end{eqnarray}
In the second line we have assumed that the dipole moment lies along the internuclear axis, i.e. $T_q^1(\delta)=\delta_{q,0}$ with the Kronecker delta $\delta_{i,j}$.
This expression can be further evaluated~\cite{brown_rotational_2003} using $\ket{\bar{\psi}}=\ket{\Omega,J,m_J}$ and $\ket{\bar{\phi}}=\ket{\Omega',J',m'_J}$ to
\begin{eqnarray}
  \mathcal{S}_\mathrm{rot}(\bar{\psi},\bar{\phi})&=&\sum_p \epsilon_p (-1)^{m_J-\Omega}\sqrt{(2J+1)(2J'+1)}\\\nonumber
  &\times&\left(
    \begin{array}{ccc}
    J & 1 & J'\\
    -m_J & p & m'_J
\end{array}
\right)
\left(
\begin{array}{ccc}
  J & 1 & J'\\
  -\Omega & 0 & \Omega'
\end{array}
\right)
  \label{Srotfinal}
\end{eqnarray}
We can identify the following selection rules from the 3-j Symbols:
\begin{itemize}
  \item $\Delta J = 0,\pm1$
  \item $\Delta \Omega = 0$
  \item $\Delta m_J=p$
\end{itemize}

\subsection{Raman transitions}\label{sec:Raman}
\begin{figure}[htpb]
  \centering
  \includegraphics{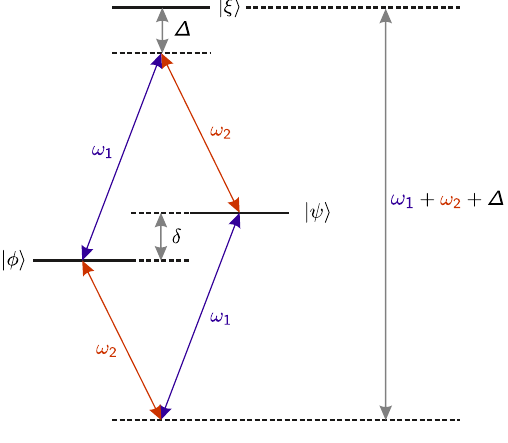}
  \caption{Relevant energy levels and frequencies for Raman transitions}
  \label{fig:RamanScheme}
\end{figure}
In a Raman transition, two states $\ket{\phi}$ and $\ket{\psi}$ with an energy gap $\delta$ are coupled via an intermediate state $\ket{\xi}$.
For that purpose two laser beams with relative detuning $\delta=\omega_1-\omega_2$  couple off-resonantly with detuning $\Delta_\xi$ to the intermediate state~(see figure~\ref{fig:RamanScheme}). The effective Rabi rate is
\begin{eqnarray}
  _\psi\Omega_\phi^\mathrm{R}&=&\frac{1}{4\hbar}\sum_\xi \frac{_\phi\Omega^{(1)}_\xi\, _\xi\Omega^{(2)}_\psi}{\Delta_\xi}+\frac{_\psi\Omega^{(1)}_\xi\, _\xi\Omega^{(2)}_\phi}{\omega_1+\omega_2+\Delta_\xi}\,.
\end{eqnarray}
The raised index for the single photon Rabi frequencies labels the associated laser beam.
In the main text, Raman transitions are used in two different contexts. Firstly, for Zeeman state preparation, where the electronic-vibrational wave function of the initial and final state are equal, i.e. $\tilde\phi=\tilde\psi$. In this context the detuning is assumed to be much larger than the energetic splittings in the excited state, therefore the detuning can be assumed to be constant and the basis for the excited state can be chosen independent of the actual angular momentum coupling case. In the case of near resonant Raman coupling the fact that the exited state follows Hund's case b) coupling has to be taken into account.

\subsubsection{ac-Stark shift}
The ac-Stark shift can be expressed as a Raman transition where $\ket{\phi}=\ket{\psi}$.
For large detuning, the difference in frequency of the individual lasers is negligible, therefore we assume $\omega=\omega_1=\omega_2$.
We find
\begin{eqnarray}
  \Delta E_\mathrm{ac}&=&\hbar _\phi\left( _\phi\Omega_\phi^\mathrm{R}(\epsilon_1)+_\phi\Omega^\mathrm{R}_\phi(\epsilon_2)\right)\\
  &=& \frac{1}{4\hbar}\sum_{\xi,n}\frac{\left(_\phi\Omega_\xi^{(n)}\right)^2}{\Delta_{\xi,n}}+ \frac{\left(_\phi\Omega_\xi^{(n)}\right)^2}{2\omega+\Delta_{\xi,n}}\\
  &=& \frac{1}{4\hbar}\sum_{\xi,n}\left(_\phi\Omega_\xi^{(n)}\right)^2\frac{2\omega_0}{\omega_0^2-\omega^2}
  \label{eq:AC-Stark}
\end{eqnarray}
where $n$ is either $1$ or $2$ and denotes the individual Raman laser beams and $\omega_0=\Delta+\omega$ is the electronic transition frequency.

Choosing the polarization of the individual Raman beams to be purely $\pi$ in one beam and $\sigma$ in the other, the ac-Stark shift can be made independent of $m_J$ by having twice as much intensity in the $\sigma$-beam compared to the $\pi$-beam.
The resulting ac-Stark shift for large detuning is independent of the populated $m_J$-state and reads
\begin{equation}
  \Delta E_\mathrm{ac}=\sum_{\tilde\xi}|\mathcal{S}_\mathrm{ev}(\tilde\phi,\tilde \xi)E_\pi|^2\frac{\omega_0}{\omega^2-\omega_0^2}
  \label{eq:AC-Stark1}
\end{equation}

\subsection{Quadrupole transition moments}
Analog to the examination of the single photon dipole transition, we can also infer a transition rate for quadrupole excitations in \osix.
The Rabi frequency associated with a quadrupole transition is given by
\begin{equation}
  _\psi\Omega_\phi^\mathrm{q}=\sum_p \frac{\Bra{\psi}T^2_p(\nabla E)\cdot T_p^2(Q(R))\Ket{\phi}}{\hbar}.
  \label{eq:QuadRabi}
\end{equation}
Considering a plane wave and assuming that the quadrupole moment lies along the internuclear axis, we can write $T^2_p(\nabla E)= |E||k|T^2_p(\epsilon)$ and $T^2_p(Q(R))=D^{(2)}_{0p}(\omega)^* T_0^2(Q(R))=D^{(2)}_{0p}(\omega)^* |Q(R)|$ and find
\begin{equation}
  _\psi\Omega_\phi^\mathrm{q}= \frac{1}{\hbar}\sum_{p=-2}^2 \mathcal{S}^{\mathrm{q}}_\mathrm{ev} \mathcal{S}^{\mathrm{q}}_{\mathrm{rot}}
  \label{eq:QuadRabi2}
\end{equation}
with
\begin{eqnarray}
  \mathcal{S}^{\mathrm{q}}_{\mathrm{ev}}&=&|k||E|\Bra{\tilde\psi}Q(R)\Ket{\tilde\phi}\\
  \mathcal{S}^{\mathrm{q}}_{\mathrm{rot}}&=& \sum_p \epsilon_p \Bra{\bar \psi}D_{0p}^{(2)}\left( \omega \right)^*\Ket{\bar \phi}
  \label{eq:SevSrot}
\end{eqnarray}
For evaluation of the electronic-vibrational part of the transition quadrupole moment, $\mathcal{S}^{\mathrm{q}}_{\mathrm{ev}}=\int_0^\infty \chi_\nu^* Q(R)\chi_{\nu'}\mathrm{d}R $, the quadrupole moment of the ground state as a function of the internuclear distance was taken from reference~\cite{feher_ab_1996} and the vibrational wave functions from Carollo \textit{et al.}~\cite{carollo_two-photon_2018}. To evaluate the integral for the quadrupole moment, we interpolated the given values by a fifth-order polynomial fit.
The rotational part of the quadrupole transition moment $\mathcal{S}^{\mathrm{q}}_{\mathrm{rot}}$ is given by
\begin{eqnarray}
  \mathcal{S}^{\mathrm{q}}_{\mathrm{rot}}&=&\sum_p \epsilon_p (-1)^{m_J-\Omega}\sqrt{(2J+1)(2J'+1)}\\\nonumber
&\times&\left(
  \begin{array}{ccc}
    J & 2  & J' \\
    -\Omega & 0 & \Omega' 
  \end{array}
\right)
  \left(
  \begin{array}{ccc}
    J & 2  & J' \\
    -m_J & p & m_J' 
  \end{array}
\right).
  \label{eq:quadRotPart}
\end{eqnarray}
The resulting selection rules are
\begin{itemize}
  \item  $\Delta J = 0, \pm 1, \pm 2$
  \item $\Delta \Omega = 0$
  \item $\Delta m_J=p$
\end{itemize}
\begin{figure}
\centering
\includegraphics[width=\columnwidth]{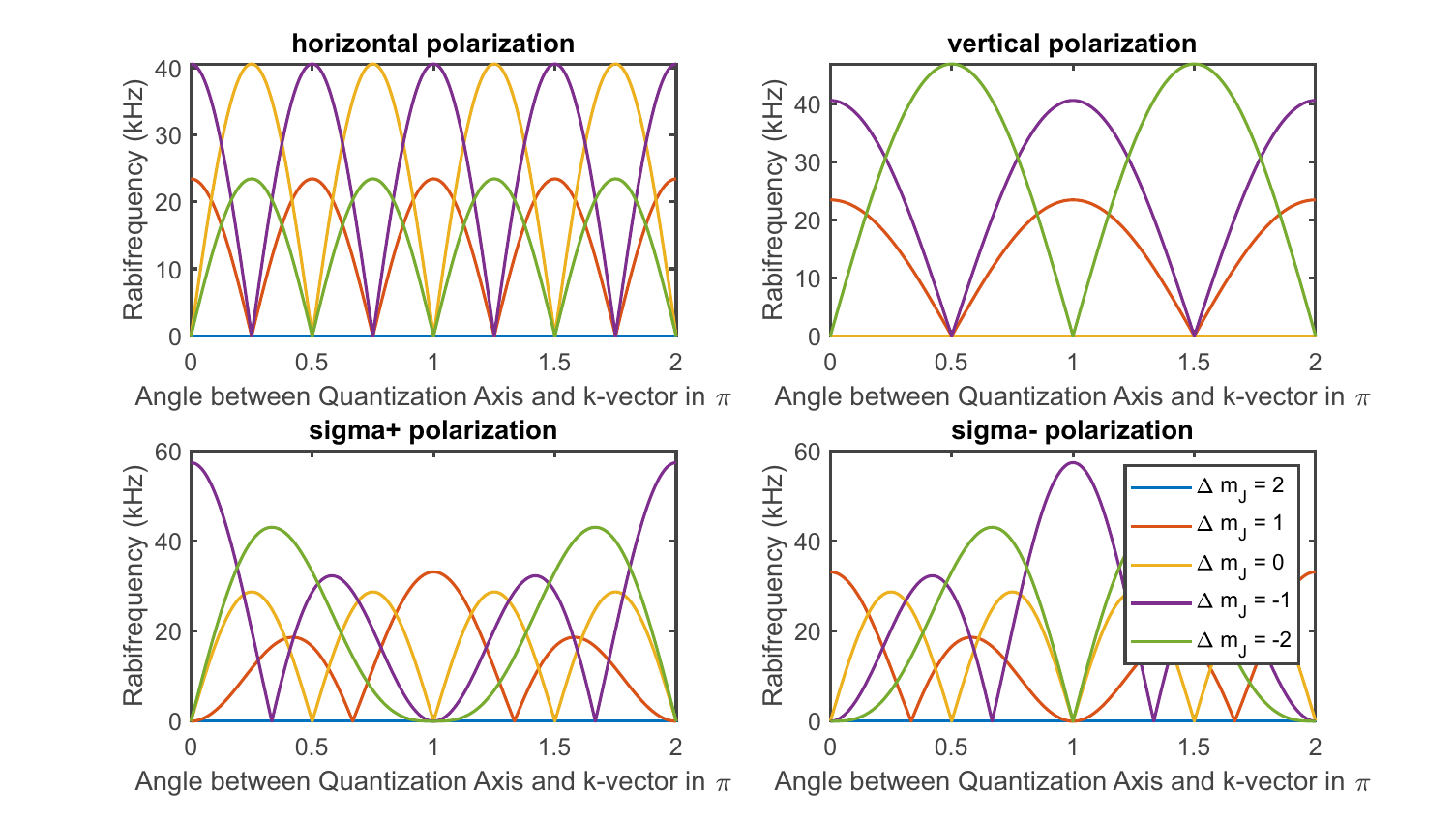}
\caption{Quadrupole Rabi excitation rate for different polarizations. The initial state is $\nu^\prime = 0, J = 1/2, m_J = 1/2$; the final state is $\nu^\prime = 1, J^\prime = 3/2, m_J + \Delta m_J$. x-axis: angle between laser direction (k-vector) and quantization axis (z-direction). y-axis: excitations per second for a laser with $1\,$W power and a waist of $10\,\upmu$m.}
\label{fig:quadEx32}
\end{figure}
\begin{figure}
\centering
\includegraphics[width=\columnwidth]{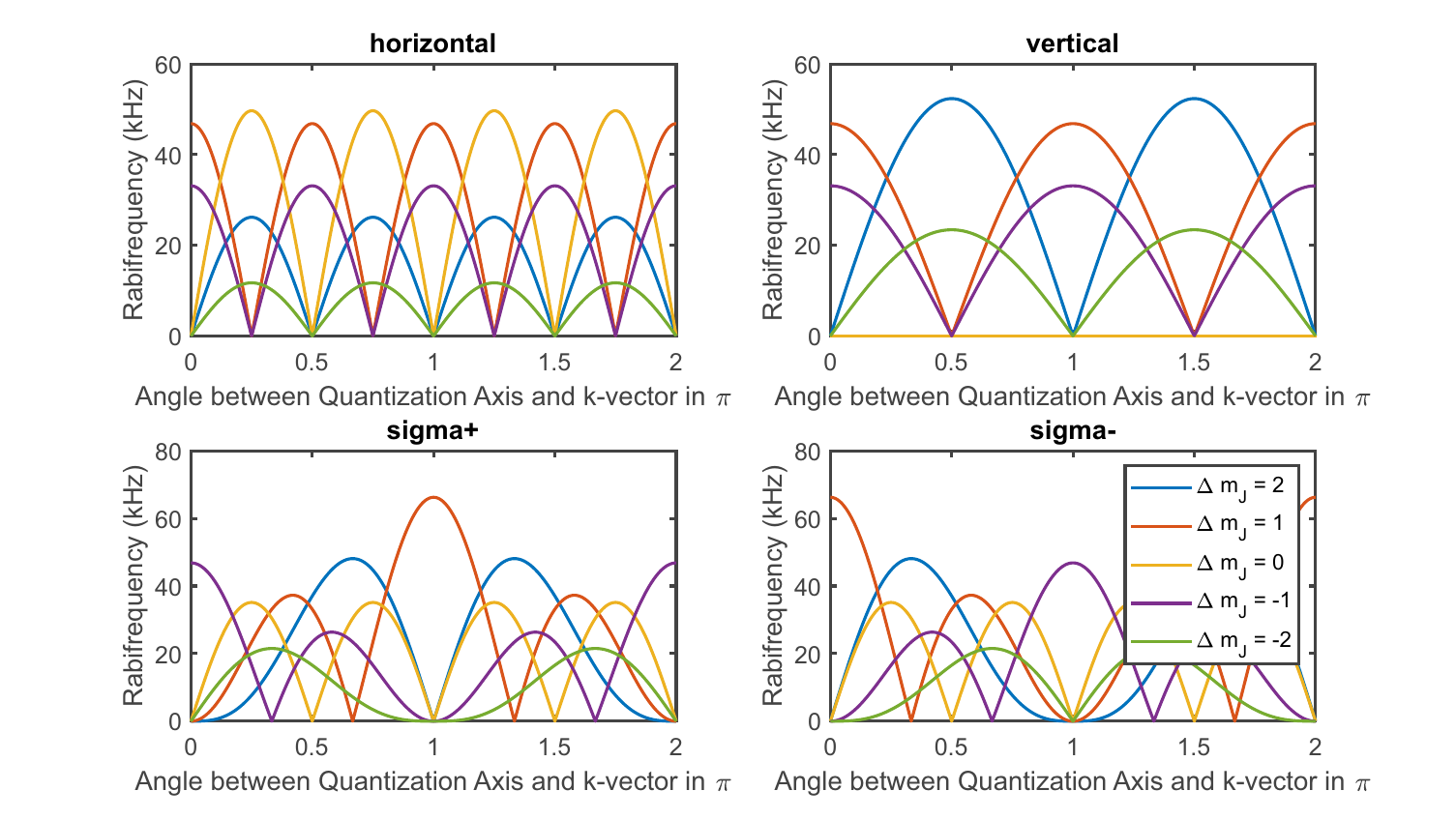}
\caption{Quadrupole Rabi excitation rate for different polarizations. The initial state is $\nu^\prime = 0, J = 1/2, m_J = 1/2$; the final state is $\nu^\prime = 1, J^\prime = 5/2, m_J + \Delta m_J$. x-axis: angle between laser direction (k-vector) and quantization axis (z-direction). y-axis: excitations per second for a laser with $1\,$W power and a waist of $10\,\upmu$m.}
\label{fig:quadEx52}
\end{figure}
\section{Schr\"odinger cat state operation}\label{sec:SC}
The state detection sequence relies on a bichromatic sideband drive that addresses the red and blue sidebands simultaneously.
In the following we will describe the dynamics induced by this interaction with a two level system.
In the interaction picture, the corresponding Hamiltonian can be written
\begin{eqnarray}
H=\ii \hbar \eta \frac{\Omega}{2} \left[\underbrace{a\sigma_+\ee^{\ii \phi_\mathrm{rsb}}}_{\mathrm{\small red- sideband}}+\underbrace{a^\dagger\sigma_+\ee^{\ii \phi_{bsb}}}_{\mathrm{\small blue-sideband}}+\mathrm{.h.c.}\right],
\end{eqnarray}
where the phases of the two light fields are given by $\phi_\mathrm{rsb}$ and $\phi_\mathrm{bsb}$. Introducing
\begin{eqnarray}
\Phi=\left(\phi_\mathrm{rsb}+\phi_\mathrm{bsb}\right)/2,\\
\delta\phi=\left(\phi_\mathrm{rsb}-\phi_\mathrm{bsb}\right)/2,
\end{eqnarray}
the Hamiltonian can be rewritten as
\begin{eqnarray}
H=\ii \hbar \eta \frac{\Omega}{2}\left(\sigma_+\ee^{\ii \Phi}+\sigma_- \ee^{-\ii \Phi }\right)\left(a\ee^{\ii \delta \phi}+a^\dagger\ee^{-\ii \delta \phi}\right)
\end{eqnarray}
By choosing $\Phi$ such that the spin rotation is around the x-axis the Hamilton operator simplifies to
\begin{eqnarray}
H=\ii \hbar \eta \frac{\Omega}{2} \sigma_x\left(a \ee^{\ii \delta \phi}+a^\dagger \ee^{-\ii \delta \phi}\right).
\label{eq:SCgeneration}
\end{eqnarray}
The effect of this Hamiltonian can be seen by expressing the associated unitary evolution operator in the $\sigma_x$ eigenbasis
\begin{eqnarray}
U_\mathrm{SC}&=&\ee^{-\frac{\ii}{\hbar}H t}\\
&=&\ket{+}\bra{+}D(\alpha)+\ket{-}\bra{-}D(-\alpha).
\end{eqnarray}
The two eigenstates of the $\sigma_x$ operator experience an oscillating force with opposite phase and their motional state is displaced in opposite directions in phase space.
Acting on the motional ground state and a spin basis state the operator creates the state
\begin{eqnarray}
\state_\mathrm{SC}=\ket{+}\ket{\alpha}+\ket{-}\ket{-\alpha},
\end{eqnarray}
which is equivalent to the Schr\"odinger cat state given in the main text. Experiments where this approach has been realized in atomic systems can be found in reference \cite{hempel_entanglement-enhanced_2013} or \cite{lo_spin-motion_2015}, for instance.
\section*{References}
 \bibliographystyle{iopart-num}

 \bibliography{OxygenProposal}

\providecommand{\newblock}{}
\begin{thebibliography}{10}
\expandafter\ifx\csname url\endcsname\relax
  \def\url#1{{\tt #1}}\fi
\expandafter\ifx\csname urlprefix\endcsname\relax\def\urlprefix{URL }\fi
\providecommand{\eprint}[2][]{\url{#2}}

\bibitem{wolf_non-destructive_2016}
Wolf F, Wan Y, Heip J~C, Gebert F, Shi C and Schmidt P~O 2016 {\em Nature\/}
  {\bf 530} 457--460 ISSN 0028-0836

\bibitem{chou_preparation_2017}
Chou C~w, Kurz C, Hume D~B, Plessow P~N, Leibrandt D~R and Leibfried D 2017
  {\em Nature\/} {\bf 545} 203--207 ISSN 0028-0836

\bibitem{sinhal_quantum-nondemolition_2020}
Sinhal M, Meir Z, Najafian K, Hegi G and Willitsch S 2020 {\em Science\/} {\bf
  367} 1213--1218 ISSN 0036-8075, 1095-9203 publisher: American Association for
  the Advancement of Science Section: Research Article
  \urlprefix\url{https://science.sciencemag.org/content/367/6483/1213}

\bibitem{chou_frequency-comb_2020}
Chou C~W, Collopy A~L, Kurz C, Lin Y, Harding M~E, Plessow P~N, Fortier T,
  Diddams S, Leibfried D and Leibrandt D~R 2020 {\em Science\/} {\bf 367}
  1458--1461 ISSN 0036-8075, 1095-9203 publisher: American Association for the
  Advancement of Science Section: Report
  \urlprefix\url{https://science.sciencemag.org/content/367/6485/1458}

\bibitem{lin_quantum_2020}
Lin Y, Leibrandt D~R, Leibfried D and Chou C~w 2020 {\em Nature\/} {\bf 581}
  273--277 ISSN 1476-4687 number: 7808 Publisher: Nature Publishing Group
  \urlprefix\url{https://www.nature.com/articles/s41586-020-2257-1}

\bibitem{collopy_effects_2023}
Collopy A~L, Schmidt J, Leibfried D, Leibrandt D~R and Chou C~W 2023 {\em
  Physical Review Letters\/} {\bf 130} 223201 publisher: American Physical
  Society
  \urlprefix\url{https://link.aps.org/doi/10.1103/PhysRevLett.130.223201}

\bibitem{brewer_27+_2019}
Brewer S~M, Chen J~S, Hankin A~M, Clements E~R, Chou C~W, Wineland D~J, Hume
  D~B and Leibrandt D~R 2019 {\em Physical Review Letters\/} {\bf 123} 033201
  \urlprefix\url{https://link.aps.org/doi/10.1103/PhysRevLett.123.033201}

\bibitem{safronova_search_2018}
Safronova M~S, Budker D, DeMille D, Kimball D~F~J, Derevianko A and Clark C~W
  2018 {\em Reviews of Modern Physics\/} {\bf 90} 025008
  \urlprefix\url{https://link.aps.org/doi/10.1103/RevModPhys.90.025008}

\bibitem{salumbides_bounds_2013}
Salumbides E~J, Koelemeij J~C~J, Komasa J, Pachucki K, Eikema K~S~E and Ubachs
  W 2013 {\em Physical Review D\/} {\bf 87} 112008 ISSN 1550-7998, 1550-2368
  \urlprefix\url{https://link.aps.org/doi/10.1103/PhysRevD.87.112008}

\bibitem{schiller_tests_2005}
Schiller S and Korobov V 2005 {\em Physical Review A\/} {\bf 71} 032505 ISSN
  1050-2947 \urlprefix\url{http://link.aps.org/doi/10.1103/PhysRevA.71.032505}

\bibitem{flambaum_enhanced_2007}
Flambaum V and Kozlov M 2007 {\em Physical Review Letters\/} {\bf 99} 150801
  ISSN 0031-9007
  \urlprefix\url{http://link.aps.org/doi/10.1103/PhysRevLett.99.150801}

\bibitem{hanneke_high_2016}
Hanneke D, Carollo R~A and Lane D~A 2016 {\em Physical Review A\/} {\bf 94}
  050101 ISSN 2469-9926, 2469-9934
  \urlprefix\url{https://link.aps.org/doi/10.1103/PhysRevA.94.050101}

\bibitem{kajita_accuracy_2017}
Kajita M 2017 {\em Physical Review A\/} {\bf 95} 023418
  \urlprefix\url{https://link.aps.org/doi/10.1103/PhysRevA.95.023418}

\bibitem{kajita_test_2014}
Kajita M, Gopakumar G, Abe M, Hada M and Keller M 2014 {\em Physical Review
  A\/} {\bf 89} 032509
  \urlprefix\url{http://link.aps.org/doi/10.1103/PhysRevA.89.032509}

\bibitem{Flambaum_dependence_2006}
Flambaum V~V and Tedesco A~F 2006 {\em Physical Review C\/} {\bf 73} 055501
  \urlprefix\url{http://link.aps.org/doi/10.1103/PhysRevC.73.055501}

\bibitem{godun_frequency_2014}
Godun R~M, Nisbet-Jones P~B~R, Jones J~M, King S~A, Johnson L~A~M, Margolis
  H~S, Szymaniec K, Lea S~N, Bongs K and Gill P 2014 {\em Physical Review
  Letters\/} {\bf 113} 210801
  \urlprefix\url{http://link.aps.org/doi/10.1103/PhysRevLett.113.210801}

\bibitem{huntemann_improved_2014}
Huntemann N, Lipphardt B, Tamm C, Gerginov V, Weyers S and Peik E 2014 {\em
  Physical Review Letters\/} {\bf 113} 210802
  \urlprefix\url{http://link.aps.org/doi/10.1103/PhysRevLett.113.210802}

\bibitem{kobayashi_measurement_2019}
Kobayashi J, Ogino A and Inouye S 2019 {\em Nature Communications\/} {\bf 10}
  3771 ISSN 2041-1723
  \urlprefix\url{http://www.nature.com/articles/s41467-019-11761-1}

\bibitem{carollo_two-photon_2018}
Carollo R, Frenett A and Hanneke D 2018 {\em Atoms\/} {\bf 7} 1 ISSN 2218-2004
  \urlprefix\url{http://www.mdpi.com/2218-2004/7/1/1}

\bibitem{mur-petit_toward_2013}
Mur-Petit J, P{\'e}rez-R{\'i}os J, Campos-Mart{\'i}nez J, Hern{\'a}ndez M~I,
  Willitsch S and Garc{\'i}a-Ripoll J~J 2013 Toward a {Molecular} {Ion} {Qubit}
  {\em Architecture and {Design} of {Molecule} {Logic} {Gates} and {Atom}
  {Circuits}\/} Advances in {Atom} and {Single} {Molecule} {Machines} ed
  Lorente N and Joachim C (Springer Berlin Heidelberg) pp 267--277 ISBN
  978-3-642-33136-7 978-3-642-33137-4
  \urlprefix\url{http://link.springer.com/chapter/10.1007/978-3-642-33137-4_20}

\bibitem{kortunov_protonelectron_2021}
Kortunov I~V, Alighanbari S, Hansen M~G, Giri G~S, Korobov V~I and Schiller S
  2021 {\em Nature Physics\/} {\bf 17} 569--573 ISSN 1745-2481 number: 5
  Publisher: Nature Publishing Group
  \urlprefix\url{https://www.nature.com/articles/s41567-020-01150-7}

\bibitem{patra_protonelectron_2017}
Patra S, Karr J~P, Hilico L, Germann M, Korobov V~I and Koelemeij J~C~J 2017
  {\em Journal of Physics B: Atomic, Molecular and Optical Physics\/} {\bf 51}
  024003 ISSN 0953-4075
  \urlprefix\url{https://doi.org/10.1088%2F1361-6455%2Faa9b92}

\bibitem{alighanbari_precise_2020}
Alighanbari S, Giri G~S, Constantin F~L, Korobov V~I and Schiller S 2020 {\em
  Nature\/} {\bf 581} 152--158 ISSN 1476-4687 number: 7807 Publisher: Nature
  Publishing Group
  \urlprefix\url{https://www.nature.com/articles/s41586-020-2261-5}

\bibitem{schmidt_spectroscopy_2006}
Schmidt P~O, Rosenband T, Koelemeij J~C~J, Hume D~B, Itano W~M, Bergquist J~C
  and Wineland D~J 2006 Spectroscopy of atomic and molecular ions using quantum
  logic {\em Proceedings of {Non}-{Neutral} {Plasma} {Physics} {VI}\/} vol 862
  (Aarhus, Denmark) pp 305--312

\bibitem{vogelius_probabilistic_2006}
Vogelius I~S, Madsen L~B and Drewsen M 2006 {\em Journal of Physics B: Atomic,
  Molecular and Optical Physics\/} {\bf 39} S1259--S1265 ISSN 0953-4075
  \urlprefix\url{http://iopscience.iop.org/0953-4075/39/19/S31}

\bibitem{leibfried_quantum_2012}
Leibfried D 2012 {\em New Journal of Physics\/} {\bf 14} 023029 ISSN 1367-2630
  \urlprefix\url{http://iopscience.iop.org/1367-2630/14/2/023029}

\bibitem{ding_quantum_2012}
Ding S and Matsukevich D~N 2012 {\em New Journal of Physics\/} {\bf 14} 023028
  ISSN 1367-2630
  \urlprefix\url{http://iopscience.iop.org/1367-2630/14/2/023028}

\bibitem{mur-petit_temperature-independent_2012}
Mur-Petit J, Garc{\'i}a-Ripoll J~J, P{\'e}rez-R{\'i}os J, Campos-Mart{\'i}nez
  J, Hern{\'a}ndez M~I and Willitsch S 2012 {\em Physical Review A\/} {\bf 85}
  022308 \urlprefix\url{http://link.aps.org/doi/10.1103/PhysRevA.85.022308}

\bibitem{shi_microwave_2013}
Shi M, Herskind P~F, Drewsen M and Chuang I~L 2013 {\em New Journal of
  Physics\/} {\bf 15} 113019 ISSN 1367-2630
  \urlprefix\url{http://iopscience.iop.org/1367-2630/15/11/113019}

\bibitem{nier_redetermination_1950}
Nier A~O 1950 {\em Physical Review\/} {\bf 77} 789--793 ISSN 0031-899X
  \urlprefix\url{https://link.aps.org/doi/10.1103/PhysRev.77.789}

\bibitem{coxon_rotational_1984}
Coxon J and Haley M 1984 {\em Journal of Molecular Spectroscopy\/} {\bf 108}
  119--136 ISSN 00222852
  \urlprefix\url{https://linkinghub.elsevier.com/retrieve/pii/002228528490290X}

\bibitem{krupenie_spectrum_1972}
Krupenie P~H 1972 {\em Journal of Physical and Chemical Reference Data\/} {\bf
  1} 423--534 ISSN 0047-2689, 1529-7845
  \urlprefix\url{http://aip.scitation.org/doi/10.1063/1.3253101}

\bibitem{brown_rotational_2003}
Brown J~M and Carrington A 2003 {\em Rotational {Spectroscopy} of {Diatomic}
  {Molecules}\/} (Cambridge University Press) ISBN 978-0-521-53078-1

\bibitem{prasad_fourier_1997}
Prasad B~C~V~V, Lacombe D, Walker K, Kong W, Bernath P and Hepburn J 1997 {\em
  Molecular Physics\/} {\bf 91} 1059--1074 ISSN 0026-8976, 1362-3028
  \urlprefix\url{http://www.tandfonline.com/doi/abs/10.1080/002689797170806}

\bibitem{liu_accurate_2015}
Liu H, Shi D, Sun J and Zhu Z 2015 {\em Molecular Physics\/} {\bf 113} 120--136
  ISSN 0026-8976, 1362-3028
  \urlprefix\url{http://www.tandfonline.com/doi/abs/10.1080/00268976.2014.948516}

\bibitem{song_rotationally_1999}
Song Y, Evans M, Ng C~Y, Hsu C~W and Jarvis G~K 1999 {\em The Journal of
  Chemical Physics\/} {\bf 111} 1905--1916 ISSN 0021-9606, 1089-7690
  \urlprefix\url{http://aip.scitation.org/doi/10.1063/1.479459}

\bibitem{hinkley__1972}
Hinkley R~K, Hall J~A, Walker T~E~H and Richards W~G 1972  {\bf 5} 204--212
  ISSN 0022-3700
  \urlprefix\url{https://doi.org/10.1088%2F0022-3700%2F5%2F2%2F016}

\bibitem{watson_isotope_1980}
Watson J~K 1980 {\em Journal of Molecular Spectroscopy\/} {\bf 80} 411--421
  ISSN 00222852
  \urlprefix\url{https://linkinghub.elsevier.com/retrieve/pii/0022285280901526}

\bibitem{tiemann_isotope_1982}
Tiemann E 1982 {\em Journal of Molecular Spectroscopy\/} {\bf 91} 60--71 ISSN
  00222852
  \urlprefix\url{https://linkinghub.elsevier.com/retrieve/pii/0022285282900303}

\bibitem{nagourney_shelved_1986}
Nagourney W, Sandberg J and Dehmelt H 1986 {\em Physical Review Letters\/} {\bf
  56} 2797--2799
  \urlprefix\url{http://link.aps.org/doi/10.1103/PhysRevLett.56.2797}

\bibitem{wan_efficient_2015}
Wan Y, Gebert F, Wolf F and Schmidt P~O 2015 {\em Physical Review A\/} {\bf 91}
  043425 \urlprefix\url{http://link.aps.org/doi/10.1103/PhysRevA.91.043425}

\bibitem{rugango_sympathetic_2015}
Rugango R, Goeders J~E, Dixon T~H, Gray J~M, Khanyile N~B, Shu G, Clark R~J and
  Brown K~R 2015 {\em New Journal of Physics\/} {\bf 17} 035009 ISSN 1367-2630
  \urlprefix\url{http://stacks.iop.org/1367-2630/17/i=3/a=035009}

\bibitem{poulsen_sideband_2011}
Poulsen G 2011 {\em Sideband {Cooling} of {Atomic} and {Molecular} {Ions}\/}
  {PhD} {Thesis} University of Aarhus Aarhus, Denmark

\bibitem{monroe_resolved-sideband_1995}
Monroe C, Meekhof D~M, King B~E, Jefferts S~R, Itano W~M, Wineland D~J and
  Gould P 1995 {\em Physical Review Letters\/} {\bf 75} 4011--4014 ISSN
  1079-7114
  \urlprefix\url{https://journals.aps.org/prl/abstract/10.1103/PhysRevLett.75.4011}

\bibitem{gebert_detection_2016}
Gebert F, Wan Y, Wolf F, Heip J~C and Schmidt P~O 2016 {\em New Journal of
  Physics\/} {\bf 18} 013037 ISSN 1367-2630
  \urlprefix\url{http://stacks.iop.org/1367-2630/18/i=1/a=013037}

\bibitem{wolf_motional_2019}
Wolf F, Shi C, Heip J~C, Gessner M, Pezz{\`e} L, Smerzi A, Schulte M, Hammerer
  K and Schmidt P~O 2019 {\em Nature Communications\/} {\bf 10} 2929 ISSN
  2041-1723 \urlprefix\url{https://www.nature.com/articles/s41467-019-10576-4}

\bibitem{ohira_phonon-number-resolving_2019}
Ohira R, Mukaiyama T and Toyoda K 2019 {\em Physical Review A\/} {\bf 100}
  060301 ISSN 2469-9926, 2469-9934
  \urlprefix\url{https://link.aps.org/doi/10.1103/PhysRevA.100.060301}

\bibitem{hendricks_all-optical_2007}
Hendricks R, Grant D, Herskind P, Dantan A and Drewsen M 2007 {\em Applied
  Physics B\/} {\bf 88} 507--513 ISSN 0946-2171
  \urlprefix\url{http://www.springerlink.com/content/1k76j29052377kmh/}

\bibitem{chen_ticking_2017}
Chen J~S 2017 {\em Ticking near the {Zero}-{Point} {Energy}: {Towards} 1 x
  10\${\textasciicircum}-\{18\}\$ {Accuracy} in {Al}\${\textasciicircum}+\$
  {Optical} {Clocks}\/} Ph.D. thesis University of Colorado at Boulder Boulder

\bibitem{sheridan_all-optical_2011}
Sheridan K, Lange W and Keller M 2011 {\em Applied Physics B\/} {\bf 104}
  755--761 ISSN 0946-2171, 1432-0649 00000
  \urlprefix\url{http://www.springerlink.com/content/523056271602k352/}

\bibitem{hannig_towards_2019}
Hannig S, Pelzer L, Scharnhorst N, Kramer J, Stepanova M, Xu Z~T, Spethmann N,
  Leroux I~D, Mehlst{\"a}ubler T~E and Schmidt P~O 2019 {\em Review of
  Scientific Instruments\/} {\bf 90} 053204 ISSN 0034-6748
  \urlprefix\url{https://aip.scitation.org/doi/10.1063/1.5090583}

\bibitem{heip_ionization_2019}
Heip J~C, Wolf F, Zawierucha M~J and Schmidt P~O 2019 {\em in preparation\/}

\bibitem{sur_optical_1985}
Sur A, Ramana C~V and Colson S~D 1985 {\em The Journal of Chemical Physics\/}
  {\bf 83} 904--905 ISSN 0021-9606, 1089-7690
  \urlprefix\url{http://aip.scitation.org/doi/10.1063/1.449506}

\bibitem{dochain_production_2015}
Dochain A and Urbain X 2015 {\em EPJ Web of Conferences\/} {\bf 84} 05001 ISSN
  2100-014X
  \urlprefix\url{https://www.epj-conferences.org/articles/epjconf/abs/2015/03/epjconf-dr2013_05001/epjconf-dr2013_05001.html}

\bibitem{gardner_multi-photon_2019}
Gardner A, Softley T and Keller M 2019 {\em Scientific Reports\/} {\bf 9} 506
  ISSN 2045-2322
  \urlprefix\url{http://www.nature.com/articles/s41598-018-36783-5}

\bibitem{tong_sympathetic_2010}
Tong X, Winney A and Willitsch S 2010 {\em Physical Review Letters\/} {\bf 105}
  143001 ISSN 0031-9007
  \urlprefix\url{http://link.aps.org/doi/10.1103/PhysRevLett.105.143001}

\bibitem{tong_collisional_2011}
Tong X, Wild D and Willitsch S 2011 {\em Physical Review A\/} {\bf 83} 023415
  \urlprefix\url{http://link.aps.org/doi/10.1103/PhysRevA.83.023415}

\bibitem{solaro_direct_2018}
Solaro C, Meyer S, Fisher K, DePalatis M~V and Drewsen M 2018 {\em Physical
  Review Letters\/} {\bf 120} 253601
  \urlprefix\url{https://link.aps.org/doi/10.1103/PhysRevLett.120.253601}

\bibitem{lien_broadband_2014}
Lien C~Y, Seck C~M, Lin Y~W, Nguyen J~H~V, Tabor D~A and Odom B~C 2014 {\em
  Nature Communications\/} {\bf 5} 4783
  \urlprefix\url{http://www.nature.com/ncomms/2014/140902/ncomms5783/full/ncomms5783.html}

\bibitem{schneider_all-optical_2010}
Schneider T, Roth B, Duncker H, Ernsting I and Schiller S 2010 {\em Nature
  Physics\/} {\bf 6} 275--278 ISSN 1745-2473
  \urlprefix\url{http://www.nature.com/doifinder/10.1038/nphys1605}

\bibitem{staanum_rotational_2010}
Staanum P~F, H{\o}jbjerre K, Skyt P~S, Hansen A~K and Drewsen M 2010 {\em
  Nature Physics\/} {\bf 6} 271--274 ISSN 1745-2473
  \urlprefix\url{http://www.nature.com/doifinder/10.1038/nphys1604}

\bibitem{ozeri_hyperfine_2005}
Ozeri R, Langer C, Jost J, DeMarco B, Ben-Kish A, Blakestad B, Britton J,
  Chiaverini J, Itano W, Hume D, Leibfried D, Rosenband T, Schmidt P and
  Wineland D 2005 {\em Physical Review Letters\/} {\bf 95} 030403 ISSN
  0031-9007
  \urlprefix\url{http://link.aps.org/doi/10.1103/PhysRevLett.95.030403}

\bibitem{schmidt_spectroscopy_2005}
Schmidt P~O, Rosenband T, Langer C, Itano W~M, Bergquist J~C and Wineland D~J
  2005 {\em Science\/} {\bf 309} 749--752
  \urlprefix\url{http://www.sciencemag.org/content/309/5735/749.abstract}

\bibitem{shaniv_atomic_2016}
Shaniv R, Akerman N and Ozeri R 2016 {\em Physical Review Letters\/} {\bf 116}
  140801 \urlprefix\url{http://link.aps.org/doi/10.1103/PhysRevLett.116.140801}

\bibitem{gebert_corrigendum_2018}
Gebert F, Wan Y, Wolf F, Heip J~C and Schmidt P~O 2018 {\em New Journal of
  Physics\/} {\bf 20} 029501 ISSN 1367-2630
  \urlprefix\url{http://stacks.iop.org/1367-2630/20/i=2/a=029501}

\bibitem{johansson_qutip_2013}
Johansson J, Nation P and Nori F 2013 {\em Computer Physics Communications\/}
  {\bf 184} 1234--1240 ISSN 0010-4655
  \urlprefix\url{http://www.sciencedirect.com/science/article/pii/S0010465512003955}

\bibitem{burd_quantum_2019}
Burd S~C, Srinivas R, Bollinger J~J, Wilson A~C, Wineland D~J, Leibfried D,
  Slichter D~H and Allcock D~T~C 2019 {\em Science\/} {\bf 364} 1163--1165 ISSN
  0036-8075, 1095-9203
  \urlprefix\url{https://science.sciencemag.org/content/364/6446/1163}

\bibitem{hempel_entanglement-enhanced_2013}
Hempel C, Lanyon B~P, Jurcevic P, Gerritsma R, Blatt R and Roos C~F 2013 {\em
  Nature Photonics\/} {\bf 7} 630--633 ISSN 1749-4885
  \urlprefix\url{http://www.nature.com/nphoton/journal/v7/n8/full/nphoton.2013.172.html?WT.ec_id=NPHOTON-201308}

\bibitem{germann_observation_2014}
Germann M, Tong X and Willitsch S 2014 {\em Nature Physics\/} {\bf 10} 820--824
  ISSN 1745-2473

\bibitem{ni_high_2008}
Ni K~K, Ospelkaus S, de~Miranda M~H~G, Pe'er A, Neyenhuis B, Zirbel J~J,
  Kotochigova S, Julienne P~S, Jin D~S and Ye J 2008 {\em Science\/} {\bf 322}
  231--235 ISSN 0036-8075
  \urlprefix\url{http://www.sciencemag.org/cgi/doi/10.1126/science.1163861}

\bibitem{ospelkaus_efficient_2008}
Ospelkaus S, Pe{\textquoteright}er A, Ni K~K, Zirbel J~J, Neyenhuis B,
  Kotochigova S, Julienne P~S, Ye J and Jin D~S 2008 {\em Nature Physics\/}
  {\bf 4} 622--626 ISSN 1745-2473, 1745-2481

\bibitem{kondov_molecular_2019}
Kondov S~S, Lee C~H, Leung K~H, Liedl C, Majewska I, Moszynski R and Zelevinsky
  T 2019 {\em Nature Physics\/} {\bf 15} 1118--1122 ISSN 1745-2481
  \urlprefix\url{https://www.nature.com/articles/s41567-019-0632-3}

\bibitem{dorfler_rotational-state-changing_2020}
D{\"o}rfler A~D, Yurtsever E, Villarreal P, Gonz{\'a}lez-Lezana T, Gianturco
  F~A and Willitsch S 2020 {\em Physical Review A\/} {\bf 101} 012706 ISSN
  2469-9926, 2469-9934
  \urlprefix\url{https://link.aps.org/doi/10.1103/PhysRevA.101.012706}

\bibitem{stenger_ultraprecise_2002}
Stenger J, Schnatz H, Tamm C and Telle H 2002 {\em Physical Review Letters\/}
  {\bf 88} 073601 ISSN 0031-9007
  \urlprefix\url{http://link.aps.org/doi/10.1103/PhysRevLett.88.073601}

\bibitem{kajita_evaluation_2016}
Kajita M 2016 {\em Applied Physics B\/} {\bf 122} ISSN 0946-2171, 1432-0649
  \urlprefix\url{http://link.springer.com/10.1007/s00340-016-6479-8}

\bibitem{feher_ab_1996}
Feh{\'e}r M and Martin P 1996 {\em Chemical Physics Letters\/} {\bf 261} 23--27
  ISSN 00092614
  \urlprefix\url{https://linkinghub.elsevier.com/retrieve/pii/0009261496009323}

\bibitem{lo_spin-motion_2015}
Lo H~Y, Kienzler D, de~Clercq L, Marinelli M, Negnevitsky V, Keitch B~C and
  Home J~P 2015 {\em Nature\/} {\bf 521} 336--339 ISSN 0028-0836
  \urlprefix\url{http://www.nature.com/nature/journal/v521/n7552/full/nature14458.html?WT.ec_id=NATURE-20150521}

\end{thebibliography}

\end{document}